\documentclass[superscriptaddress,aps,prX,twocolumn,showpacs,nofootinbib,longbibliography,notitlepage,floatfix]{revtex4-2}
\usepackage{etex}
\usepackage{amsmath,amssymb,amsthm}
\usepackage[colorlinks=true,citecolor=blue,urlcolor=blue]{hyperref}
\usepackage[pdftex]{graphicx}
\usepackage{times,txfonts}
\usepackage{braket}
\usepackage{color}
\usepackage{natbib}
\usepackage{amsmath,blkarray}
\usepackage{mathtools}
\usepackage{ulem}
\usepackage{latexsym}
\usepackage{tabularx, booktabs}
\usepackage{graphics,epstopdf}
\usepackage{graphicx}
\usepackage{float}
\usepackage{amsfonts}
\usepackage{tikz}
\usetikzlibrary{quantikz}
\usepackage{color,soul}

\newcommand{\be}{\begin{equation}}
	\newcommand{\ee}{\end{equation}}
\newcommand{\ba}{\begin{eqnarray}}
	\newcommand{\ea}{\end{eqnarray}}

\usepackage{multirow}
\usepackage{appendix}
\usepackage{url}

\begin{document}
	
	\title{Meta-Learning Digitized-Counterdiabatic Quantum Optimization} 
	
	\author{Pranav Chandarana}
	\thanks{Co-first authors.}
	\affiliation{Department of Physical Chemistry, University of the Basque Country UPV/EHU, Apartado 644, 48080 Bilbao, Spain}
	\affiliation{EHU Quantum Center, University of the Basque Country UPV/EHU, 48940 Leioa, Spain}
	
	\author{Pablo S. Vieites}
	\thanks{Co-first authors.}
	\affiliation{Department of Physical Chemistry, University of the Basque Country UPV/EHU, Apartado 644, 48080 Bilbao, Spain}
	
	\author{Narendra N. Hegade}
	\affiliation{International Center of Quantum Artificial Intelligence for Science and Technology~(QuArtist) \\ and Physics Department, Shanghai University, 200444 Shanghai, China}
	
	\author{Enrique Solano}
	\affiliation{International Center of Quantum Artificial Intelligence for Science and Technology~(QuArtist) \\ and Physics Department, Shanghai University, 200444 Shanghai, China}
	\affiliation{IKERBASQUE, Basque Foundation for Science, Plaza Euskadi 5, 48009 Bilbao, Spain}
	\affiliation{Kipu Quantum, Kurwenalstrasse 1, 80804 Munich, Germany}
	
	\author{Yue Ban}
	\affiliation{Department of Physical Chemistry, University of the Basque Country UPV/EHU, Apartado 644, 48080 Bilbao, Spain}
	\affiliation{EHU Quantum Center, University of the Basque Country UPV/EHU, 48940 Leioa, Spain}
	\affiliation{TECNALIA, Basque Research and Technology Alliance (BRTA), 48160 Derio, Spain}
	
	\author{Xi Chen}
	\email{chenxi1979cn@gmail.com}
	\affiliation{Department of Physical Chemistry, University of the Basque Country UPV/EHU, Apartado 644, 48080 Bilbao, Spain}
	\affiliation{EHU Quantum Center, University of the Basque Country UPV/EHU, 48940 Leioa, Spain}

	\date{\today}
	
	\begin{abstract}
		Solving optimization tasks using variational quantum algorithms has emerged as a crucial application of the current noisy intermediate-scale quantum devices. However, these algorithms face several difficulties like finding suitable ansatz and  appropriate initial parameters, among others. In this work, we tackle the problem of finding suitable initial parameters for variational optimization by employing a meta-learning technique using recurrent neural networks. We investigate this technique with the recently proposed digitized-counterdiabatic quantum approximate optimization algorithm (DC-QAOA) that utilizes counterdiabatic protocols to improve the state-of-the-art QAOA. The combination of meta learning and DC-QAOA enables us to find optimal initial parameters for different models, such as MaxCut problem and the Sherrington-Kirkpatrick model. Decreasing the number of iterations of optimization as well as enhancing the performance, our protocol designs short depth circuit ansatz with optimal initial parameters by incorporating shortcuts-to-adiabaticity principles into machine learning methods for the near-term devices.
	\end{abstract}
	
	\maketitle
	
	\section{\label{intro} Introduction}
	Among the recent advancements in quantum algorithms, hybrid quantum-classical optimization has been of great importance to the quantum computing community due to its relevance for the near-term devices~\cite{peruzzo2014variational,killoran2019continuous,wecker2015progress,biamonte2017quantum,zhou2020quantum,RevModPhys.94.015004}. These variational quantum algorithms (VQAs) take advantage of quantum circuits with variational parameters to minimize a cost function with the aid of classical optimization routines. Of all the VQAs, the quantum approximate optimization algorithm (QAOA)~\cite{farhi2014quantum} has proved to be a promising algorithm to tackle combinatorial optimization problems~\cite{zhou2020quantum}. Although QAOA has been proposed to demonstrate quantum supremacy~\cite{farhi2016quantum}, there are still several challenges that need to be addressed.
	
	For any variational quantum optimzation algorithm, one of the major challenges is to increase the expressibility of the circuit ansatz so that more parts of the Hilbert space can be reached. One way to accomplish this is by adding terms to the circuit ansatz with more free parameters. As QAOA closely resembles the quantum adiabatic evolution, many works propose the use of shortcuts to adiabaticity (STA)~\cite{guery2019shortcuts,torrontegui2013shortcuts} to improve the state-of-the-art QAOA. As a toolbox of quantum control including techniques as invariant engineering~\cite{chen2010fast,chen2011lewis}, fast-forward~\cite{masuda2008fast,masuda2010fast}, counterdiabatic (CD) protocols~\cite{demirplak2003adiabatic,demirplak2005assisted,berry2009transitionless,PhysRevLett.111.100502}, STA methods were developed to circumvent the need for slow driving~\cite{takahashi2019hamiltonian}. Among them, implementing approximate counterdiabaticity in quantum many-body systems \cite{sels2017minimizing,ClaeysPRL}
	has shown drastic improvement in adiabatic computing~\cite{hegade2021shortcuts}, factorization problems~\cite{hegadePRAletter}, portfolio optimization~\cite{hegade2021portfolio}, preparing entangled state~\cite{opatrny2014partial,petiziol2019accelerating,zhou2020experimental,ji2019experimental}, and quantum annealing~\cite{passarelli2020counterdiabatic,takahashi2017shortcuts,vinci2017non}. In particular, a study solving a class of optimization problems called ``digitized counterdiabatic quantum optimization" has reported a polynomial enhancement over the current methods, in which CD interaction serves as a non-stoquastic catalyst~\cite{hegade2022digitized}. Among the algorithms using CD protocols in QAOA~\cite{yao2021reinforcement,chandarana2022digitized,wurtz2022counterdiabaticity,PhysRevA.105.042415}, the digitized-counterdiabatic quantum approximate optimization algorithm (DC-QAOA)~\cite{chandarana2022digitized} has shown drastic improvement over standard QAOA. In DC-QAOA, appropriate terms are 
	supplemented to the circuit ansatz corresponding to the adiabatic gauge potentials~\cite{sels2017minimizing} to reach the ground state more efficiently. Algorithms like these are still an active subject of research as there are wide range of possibilities of further improvement. 
	
	The second challenge is local optimization. Gradient-based optimization~\cite{schuld2019evaluating} routines make use of either quantum form of back-propagation of errors~\cite{verdon2018universal} or finite-difference gradients~\cite{farhi2018classification,chen2021universal}. Both these methods are expensive as additional circuit depth and a lot of measurements are required respectively. One of the ways to solve the challenge of local optimization is
	to set suitable initialization of parameters that are close to global minima, since, by doing so, the optimization iterations decrease significantly. Several strategies for parameter initialization have been already reported~\cite{zhou2020quantum,yang2017optimizing,grant2019initialization,Egger2021warmstartingquantum,Sack2021quantumannealing}.
	
	The incorporation of machine learning into quantum algorithms has been reported to enhance the performance, for instance, using RL algorithms~\cite{sutton2018reinforcement,rose2021reinforcement} such as Q-learning~\cite{bukov2018reinforcement1,bukov2018reinforcement2}, policy gradient~\cite{niu2019universal,august2018taking,porotti2019coherent} and Alphazero~\cite{dalgaard2020global} in VQAs, policy gradient as an alternative optimizer for QAOA~\cite{yao2020policy}, the implementation of Q-learning to solve combinatorial problems~\cite{garcia2019quantum},  
	finding the groundstate of transverse Ising model~\cite{wauters2020reinforcement} by proximal policy optimization (PPO), hybrid optimization by Monte Carlo search~\cite{yao2022monte} etc.
	
	In this article, we tackle the second challenge i.e. to enhance the classical optimization part, by using recurrent neural networks (RNNs) as a black-box optimizer to find good initial parameters for the circuit. In this technique, RNNs are allowed to learn from the behaviour of the gradients to predict the optimal initial state, known as meta-learning~\cite{chen2017learning}. Two kinds of recurrent neural networks namely Long short-term memory (LSTM) cells and Gated recurrent units (GRU) are applied to DC-QAOA. The implementation of RNN-enhanced DC-QAOA into MaxCut and SK model show that the trained RNN always finds suitable parameters which are near the ones corresponding to the exact ground state. We also compare these two RNNs in terms of performance and resources to check which of them is suitable for these type of problems.
	
	The article is structured as follows. In the next section, DC-QAOA is briefly discussed with the incorporation of RNNs as a classical optimization routine. In Sec.~\ref{results}, we show the results with MaxCut problem for unweighted and weighted MaxCut instances and SK model. Sec.~\ref{disc} is devoted to discussions and future works.
	\section{Meta-Learning DC-QAOA}
	\begin{figure}
		\centering
		\includegraphics[width=\linewidth]{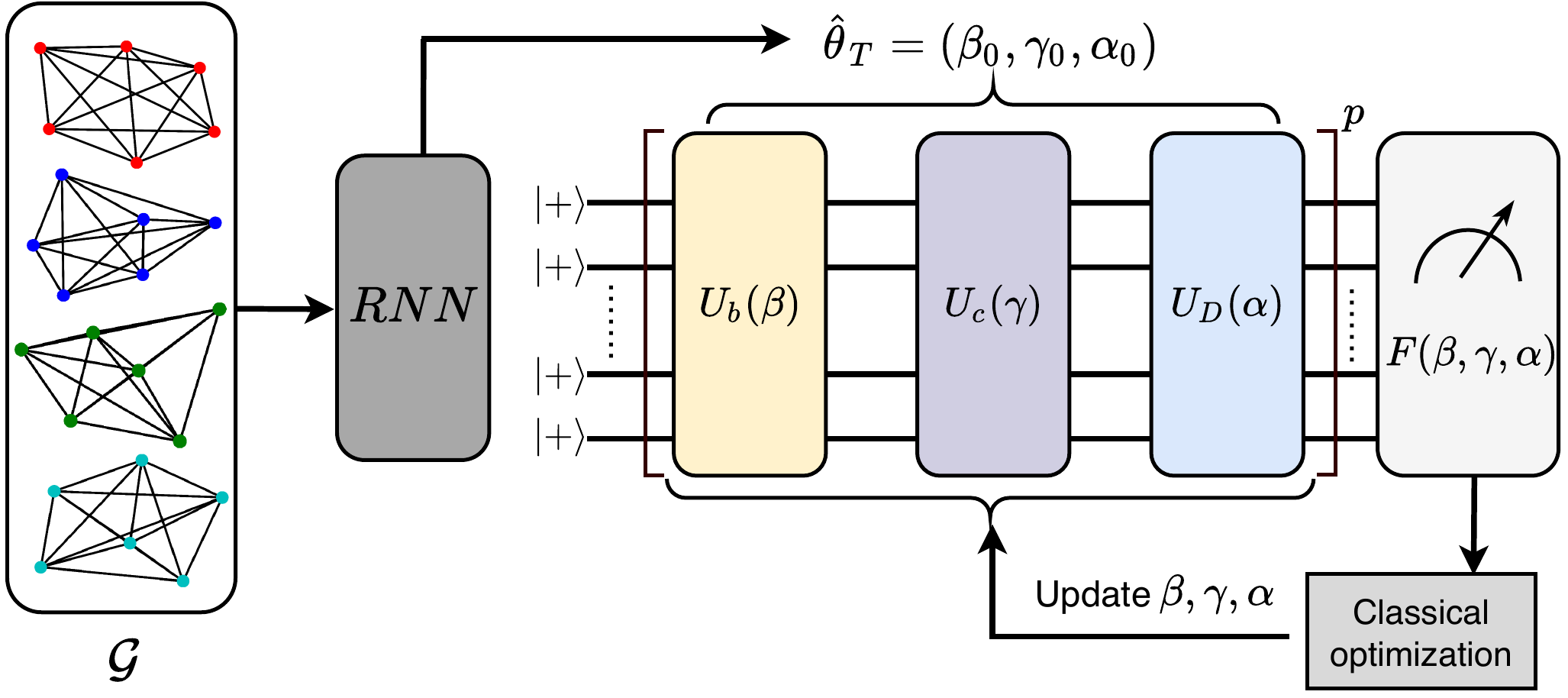}
		\caption{Schematic diagram showing the LSTM cell with a set of input graphs $\mathcal{G}$ that outputs input parameters $(\beta_0,\gamma_0,\alpha_0)$ for DC-QAOA. $F(\beta_,\gamma_,\alpha)$ is the cost function at each iteration. $p$ shows the number of layers of unitaries $U_b(\beta)$, $U_c(\gamma)$, and $U_D(\alpha)$.}
		\label{RNNcircuit}
	\end{figure}
	We commence by briefly reviewing DC-QAOA. DC-QAOA can be interpreted as the more generalized version of QAOA. It falls under the category VQAs where a quantum circuit with variational parameters is utilized along with classical optimization routines to optimize a cost function $F$. Generally, cost function $F$ corresponds to the expectation value of the cost Hamiltonian $H_c$ to the quantum state $\ket{\psi_f(\gamma,\beta)}$ given by
	\begin{equation}
		F(\gamma,\beta) =\bra{\psi_f(\gamma,\beta)}H_c\ket{\psi_f(\gamma,\beta)}\label{F}.
	\end{equation}
	
	As far as the quantum circuit is concerned, a $p$-depth QAOA circuit ansatz has an iterative application (upto $p$ times) of two unitaries: mixer term $U_b(\beta)$ and Hamiltonian term $U_c(\gamma)$ applied iteratively $p$ times to an initial state  $\ket{\psi_i}= \ket{+}^{\otimes N}$ in the computational basis. Hence, the final state $\ket{\psi_f}$ will be
	\begin{equation}
		\ket{\psi_f(\gamma,\beta)} =  U( \gamma, \beta)\ket{\psi_i},
	\end{equation}
	with
	\begin{equation*}
		U(\gamma, \beta) = U_m{(\beta_p)}U_c{(\gamma_p)}U_m{(\beta_{p-1})}U_c{(\gamma_{p-1})} \dots U_m{(\beta_1)}U_c{(\gamma_1)},
	\end{equation*}
	where $U_m(\beta)= e^{-i\beta \sum_i \sigma_i^x}$ and $U_c(\gamma)=e^{-i\gamma H_c}$. Here, $(\beta,\gamma)$ are free parameters optimized by the classical optimization routines that can be either local or global optimizations. A systematic comparison of different types of optimizers is given in Ref.~\cite{lockwood2022empirical}. The choice of QAOA ansatz is beneficial due to its scalability and good trainability~\cite{PRXQuantum.1.020319,anschuetz2021critical}. However, the high-depth QAOA may become impractical to implement experimentally due to various noise sources~\cite{weidenfeller2022scaling}. Therefore, improvising low-depth QAOA is an active area of research. To this end, many adaptations have been reported to QAOA to improve the performance~\cite{zhu2020adaptive,a12020034,headley2020approximating}. 
	In DC-QAOA, an additional term $U_D(\alpha)$ called the CD unitary is added to each layer with $U_c(\gamma)$ and $U_b(\beta)$. This inclusion has shown several advantages. First, the additional free parameter $\alpha$ makes the ansatz more expressible, and second, the CD unitary effectively decreases circuit depth $p$ to reach a target state. The information of all the spectral properties is needed to calculate the exact CD term~\cite{Demirplak2003,PhysRevLett.109.115703,berry2009transitionless} which might not be available for all the cases thus we choose the approximate CD term from the pool defined by using the nested commutator ansatz method given by~\cite{sels2017minimizing}
	\begin{equation}
		\label{gauge}
		A_{\lambda}^{(l)} = i \sum_{k = 1}^l \alpha_k(t) \underbrace{[H_{ad},[H_{ad},......[H_{ad},}_{2k-1}\partial_{\lambda} H_{ad}]]],
	\end{equation}
	where $l$ gives the order of expansion and $ H_{ad}(\lambda) = (1-\lambda) H_{mixer}+ \lambda H_c$. Keeping ourselves constrained to only two-local terms, the CD pool is given by  $A^{(2)}_{\lambda} = \{\sigma^y, \sigma^z \sigma^y, \sigma^y \sigma^z,  \sigma^x \sigma^y,  \sigma^y  \sigma^x\}$. It is worthwhile here to point out that higher order terms can also be chosen but that results in increased circuit depth so we restrict ourselves to the second order terms. The next step would be to randomly initialize the parameters, feed them into the circuit and compute $F(\beta,\gamma,\alpha)$. A classical optimizer updates parameters $(\beta,\gamma,\alpha)$ until $(\beta',\gamma',\alpha')$ such that $F(\beta',\gamma',\alpha')$ is minimized. However, randomly chosen parameters can be away from the global minima such that the classical optimization process becomes extremely difficult. To circumvent this, instead of using a classical optimizer directly, we use meta-learning by implementing a RNN which is trained to obtain optimal initial parameters, as shown in Fig.~\ref{RNNcircuit}.

	Meta-learning is a widespread technique that consists of using an algorithm whose output will be updating the parameters of some learning algorithm, in order to improve a given objective~\cite{hospedales2021}. That is why meta-learning is also known as ``learning to learn" as these algorithms learn from other algorithms being used. In this context, recurrent neural networks (RNNs) have been proven to be very useful to improve the learning algorithm~\cite{10.1162/NECO_a_00263}. In feedforward networks, an input vector ${\theta}$ (which in our case represents the angles of our variational circuit) goes through the different weight layers to get the output ${\hat {y}}$. In RNNs a hidden state $h_t$ will be introduced to keep information from previous iterations of the network, allowing the network to exhibit temporal behavior and thus having time-dependent inputs and outputs ${\theta_t}$ and ${\hat{y}_t}$. This way, the output of a RNN at time step $T$ results in a joint probability distribution,
	\begin{equation}
		\hat{y}_T=\prod_{t=1}^{T}p(\theta_t|\theta_{t-1},...,\theta_{1})\equiv g(h_T),
		\label{eq:output}
	\end{equation}
	where the function $g$ will be chosen depending on the problem we are modelling, and will have into account the network weights and biases.

\begin{figure}[t]
		\centering
		\includegraphics[width=1.0\linewidth]{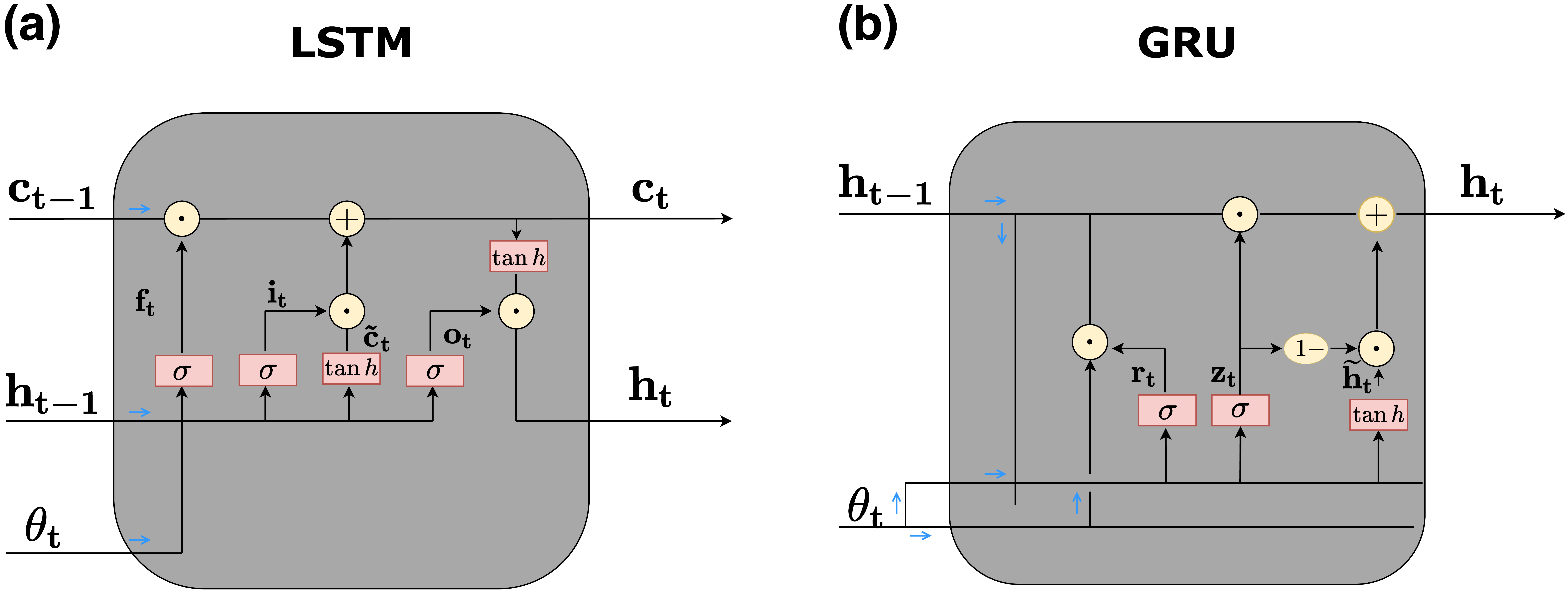}
		\caption{Schematic diagram showing information flow on LSTM and GRU cell for a time step $t$. The gates will modulate the amount of information kept from previous iterations and input vectors. $h_t$ shows the hidden state, $c_t $ shows the cell state and $\theta_t$ shows the input (in our case the optimizable parameters). It can be seen how GRU uses one less door and doesn't use the cell state $c_t$. The symbol $\odot$ denotes the Hadamard product. }
		\label{fig:LSTM_GRU}
	\end{figure}

\begin{figure}[t]
		\centering
		\includegraphics[width=1\linewidth,height=6cm]{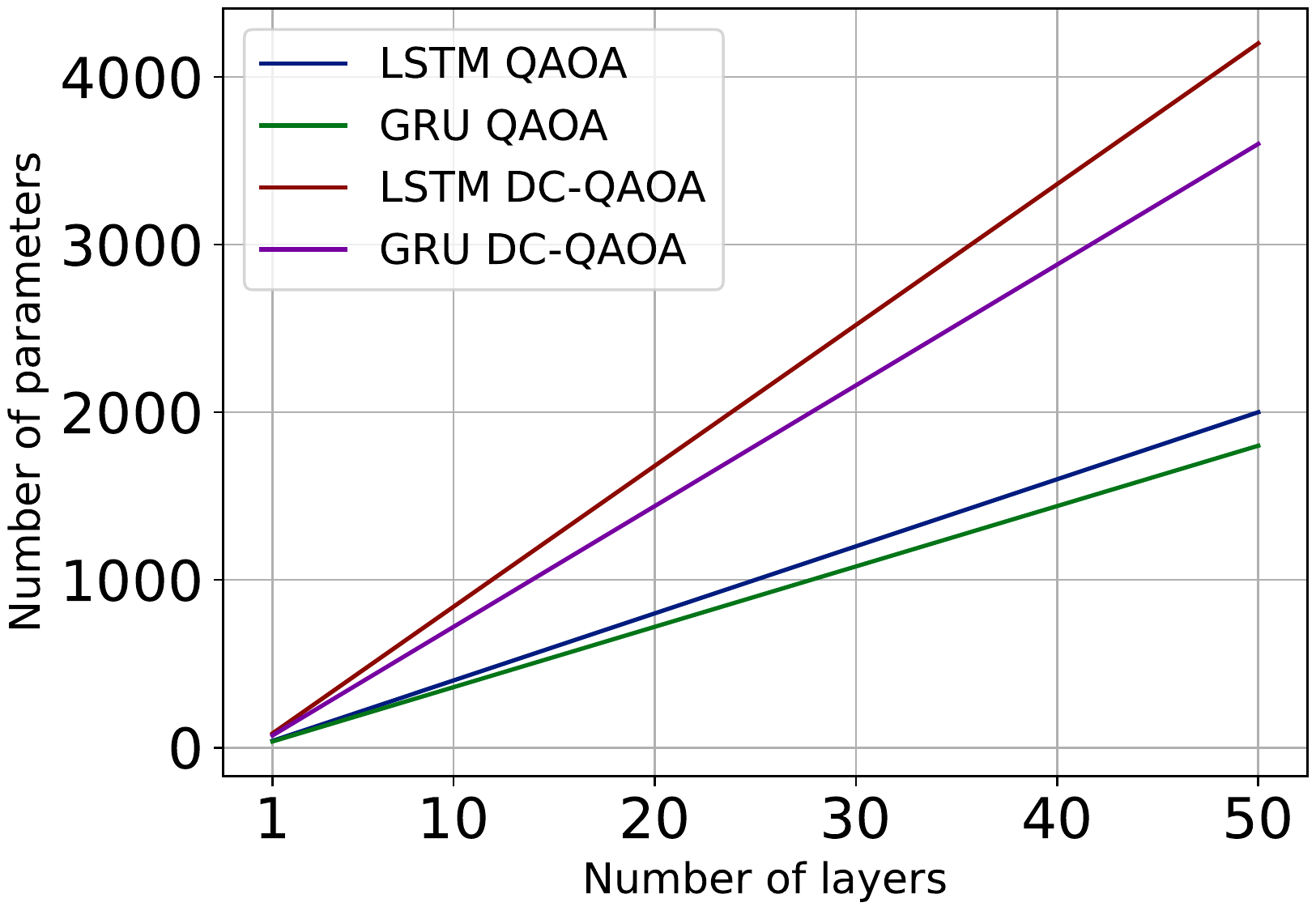}
		\caption{Trainable parameters of weight matrices and biases as a function of the number of layers for QAOA and DC-QAOA.}
		\label{fig:trainable_parameters}
	\end{figure}

	Long short term memory models (LSTMs) \cite{hochreiter} and gated recurrent Units (GRUs) \cite{cho2014properties}  are some of the most popular RNNs. As the LSTM name suggests, it works in a way such that long-term and short-term dependancies on the data can be identified. GRU is a variation of LSTM, uses fewer gate operations to serve the same purpose. An advantage of these networks is that they partially solve the problem of vanishing and exploding gradients that happen when training the networks using backpropagation methods ~\cite{lechner2020learning,info12110442}.   Fig.~\ref{fig:LSTM_GRU} shows a schematic diagram of a LSTM and GRU cells. Detailed discussion about how information flows in each of these cells is given in Appendix.~\ref{Appinfo}. One of the major differences between LSTM and GRU is the number of operations in the cell. GRU has lower number of gates so less resources will be used in contrast with LSTM. A comparison of optimizable parameter requirements for LSTM and GRU with both QAOA and DC-QAOA with the numbers of layers is shown in Fig.~\ref{fig:trainable_parameters}. It can be seen that the resources required for the DC-QAOA scale significantly faster as compared to QAOA. This is due to the fact that DC-QAOA requires 3 parameters per layer while QAOA requires 2 parameters. This prove to be a vital aspect when the scalability of the technique is in question. In the following sections, we show that GRU can be a good candidate for this technique if the algorithm is scaled to higher system sizes.

	The input vectors of the network $x_t$ will be the angles of the variational circuit $\theta_t$. On each time step the network will output the guesses of the parameters $\hat{y}_t=g(h_t)$. As our hidden state $h_t \in [-1,1]$, we choose $g(h_t)=\pi h_t=\theta_{t+1}$. This way the hidden state at step $t$ serves as input for the state $t+1$, covering all possible angles for the circuit. The loss function $\mathcal{L}(\hat{\theta})$ used to train the network (matrices of weights $W$ and biases $b$) can be formulated in different ways. In our case, we will use the following expression 
	\begin{equation}
		\mathcal{L}(\hat{\theta}) = \sum_{t=1}^{T}\omega_{i} F(\theta_t),
		\label{lossfn}
	\end{equation}
	where $T$ is the number of time steps, $F(\theta_t)$ is the same as that in Eq.~\eqref{F}, and $\omega_i$ are arbitrary weights assigned depending on how much importance we want to give to each step. Implementation of LSTM for finding initial parameters was also reported by Ref.~\cite{verdon2019learning}. In this article, we include GRU alongwith LSTM and instead of using QAOA, we use DC-QAOA with an aim to devise a more efficient and scalable technique. In the next section, we apply meta-learning to DC-QAOA for MaxCut and SK model to demonstrate that merging these two paradigms results into significant advantage in both reducing the number of iterations and getting higher success in terms of finding the ground state.
	
	\section{Results}\label{results}
	\subsection{MaxCut}
	To benchmark our method, we begin with the MaxCut problem first, by considering a graph $\mathcal{G} = (\mathcal{V},\mathcal{U})$ with $\mathcal{V}$ representing the set of vertices and $\mathcal{U}$ the set of edges. In the MaxCut problem, the task is to find a subset of vertices $\mathcal{V}_0$ and its complement $\mathcal{V}_1$ that maximize the number of edges between $\mathcal{V}_0$ and $\mathcal{V}_1$. Regarding the complexity, MaxCut is known to be a NP-complete problem ~\cite{karp1972reducibility}.  Despite the promising results of QAOA applied for MaxCut problem, it was recently proposed that it still might require hundreds of qubits to actually reach quantum advantage~\cite{Guerreschi2019}.
	
	To encode this into a Hamiltonian, we can use a bitstring $z = z_1z_2...z_V$ such that if $z_i=1$, the $i$th vertex is in set $\mathcal{V}_0$ and if $z_i=-1$, the $i$-th vertex is in set $\mathcal{V}_1$. Hence the cost Hamiltonian which we will minimize will be,
	\begin{equation}
		H_c=-\frac{1}{2} \sum_{(i, j) \in E} w_{i j}\left(1-\sigma_z^{i} \sigma_z^{j}\right),
	\end{equation}
	where $w_{ij}$ are the weights assigned to edges. To verify the performance of our method, we consider two types of problems: A 10-node unweighted 3-regular problem and a 10-node complete weighted case problem.
	
	In order to make computations with DC-QAOA we choose $U_D(\alpha) = e^{-\frac{1}{2}i\alpha \sum_{ij}w_{ij}\sigma_z^i\sigma_y^j}$, where $i,j$ run over the edges of the respective graph. To make this choice, 
	we compare the performance of various CD terms (Appendix ~\ref{appendixB}), from which we chose the best performing CD term for our simulations.
	
	We generate random graph instances of unweighted 3-regular and weighted complete problems and trained a type of RNN for each problem and variational circuit (with or without CD terms) having then a total of 4 different RNNs. Then we choose a set of 10 random graphs for each problem and optimize them with random and LSTM/GRU initialization. The time horizon (number of iterations of the network) is set to $T=6$. On each iteration of the network we introduce as inputs: the parameters obtained from previous iterations, the hidden state and the cell state in the case of LSTM cell. The cell parameters are initialized to 0 every time we start the cells, and the initial angles of the variational circuit are near to 0. The loss function used is given by Eq.~\eqref{lossfn} with the parameters $\theta$s depending on the algorithm used, i.e., $(\gamma,\beta)$ for QAOA and $(\gamma,\beta,\alpha)$ for DC-QAOA. To obtain the gradients for training the network, we use Adam optimizer~\cite{kingma2014adam} for the unweighted case and Adagrad optimizer ~\cite{JMLR:v12:duchi11a} for the weighted one, both with a learning rate of $0.1$. The number of graphs trained for unweighted case are $100$ and for weighted case $300$ graphs were trained. The maximum number of epochs is chosen to be $10$. However, we introduce a tolerance $\tau=0.01$ such that if $|\mathcal{L}^{epoch}_{t+1}-\mathcal{L}^{epoch}_{t}| \leq 0.01$, then the training process is stopped. To quantify the performance,  we define the relative error $\mathcal{E}$ given by,
	\begin{equation}
		\mathcal{E} = \frac{F(\beta,\gamma,\alpha) - E_0}{|E_0|} ,
	\end{equation}
	\begin{figure}
		\centering
		\includegraphics[width=1\linewidth]{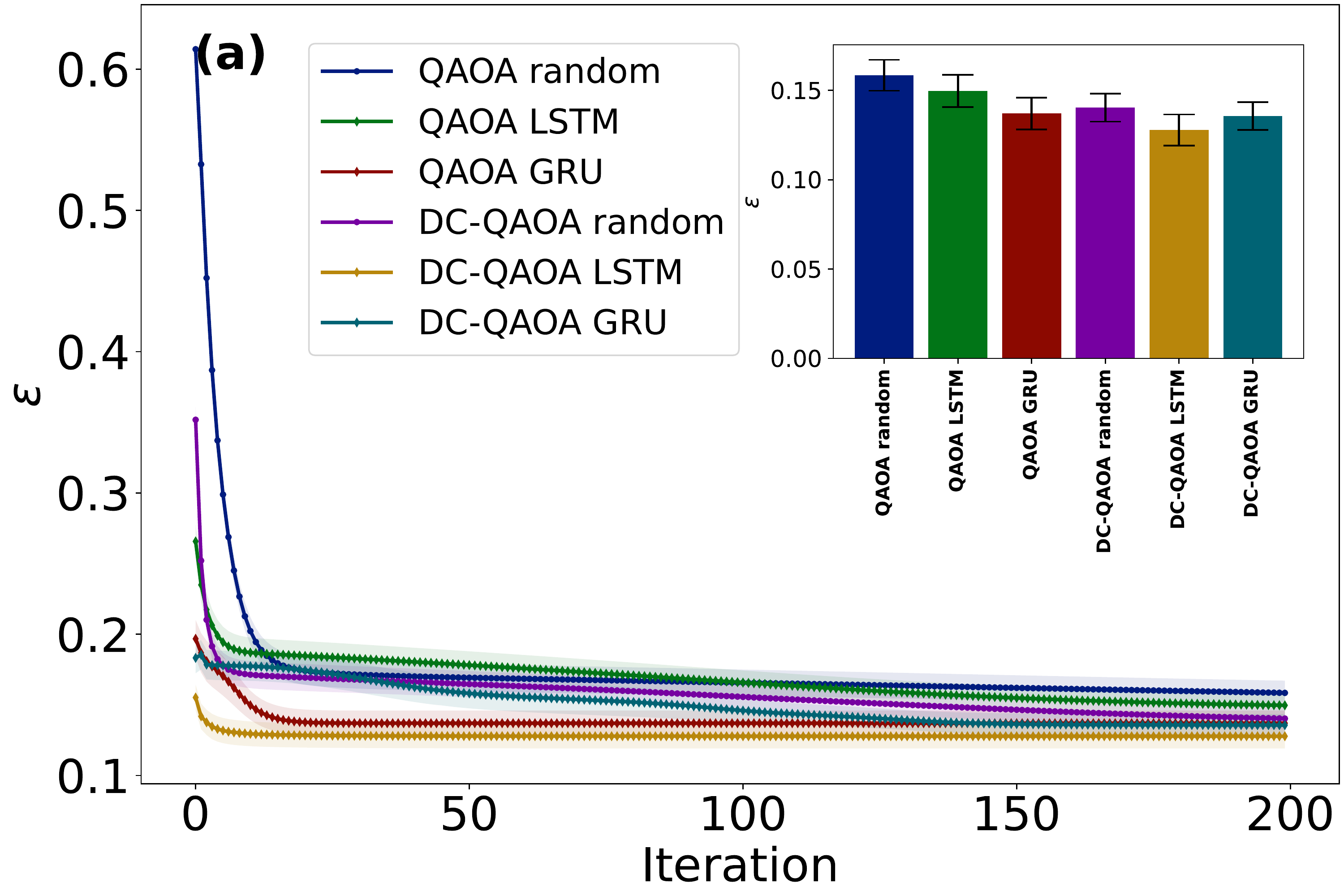}
		\includegraphics[width=1\linewidth]{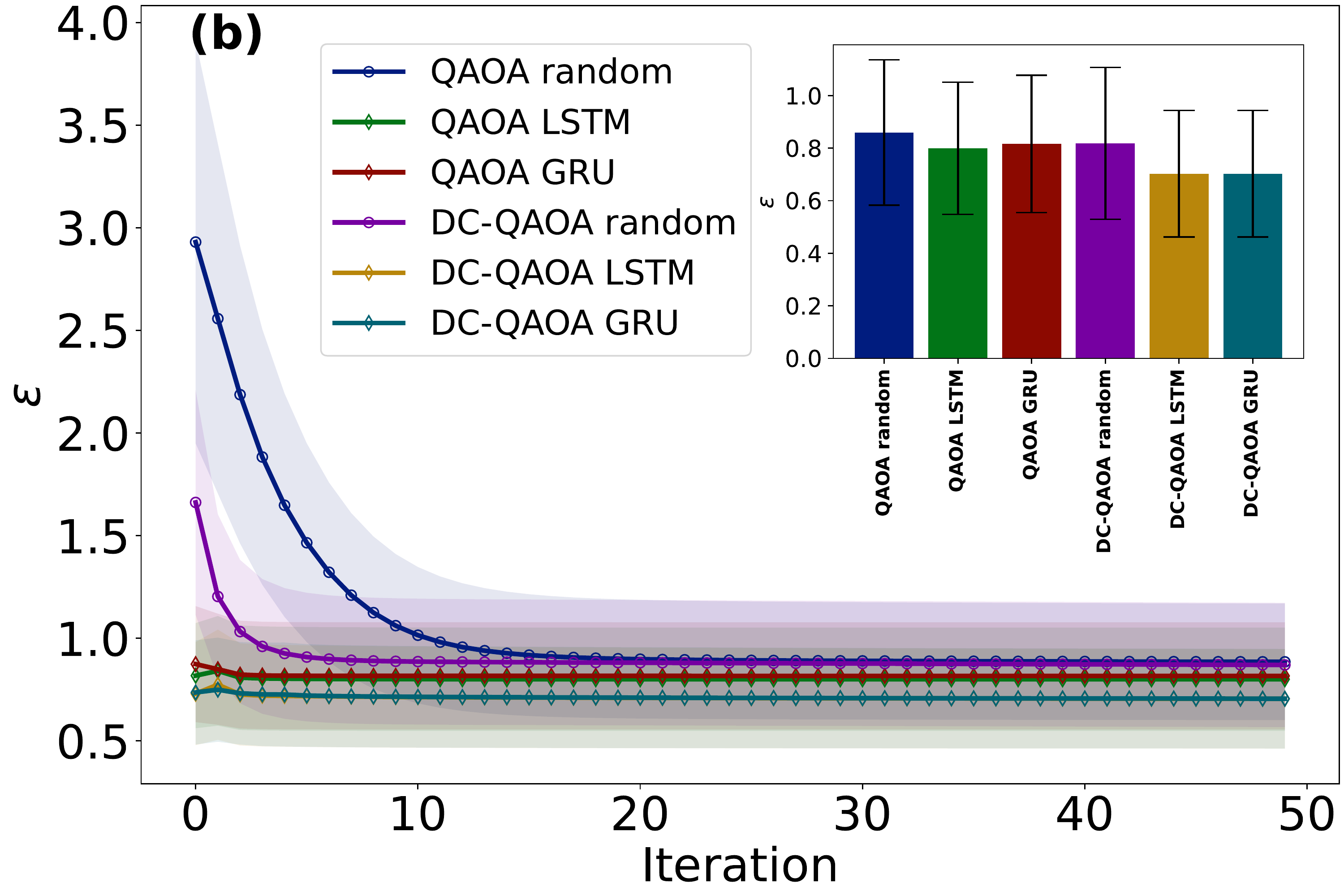}
		\label{fig:CD_weighted}
		\caption{Relative error ($\mathcal{E}$) as a function of number of iterations for (a) 3-regular unweighted MaxCut and (b) first 50 steps of complete weighted MaxCut problem. Different colored line compare results of QAOA and DC-QAOA  for $p=2$ with randomly initialized parameters and with LSTM/GRU initialized parameters. The bar plots in the corner show the results of last step of optimization. Shaded region show standard error. Results were obtained on an ideal quantum simulator.}
		\label{fig:MaxCut}
	\end{figure}
	where $E_0$ is the energy of the ground state configuration.
	
	In Fig.~\ref{fig:MaxCut}, $\mathcal{E}$  as the function of number of iterations is shown for 10 qubit unweighted and weighted MaxCut problem to compare DC-QAOA and QAOA for $p=2$ with randomly initialized parameters and LSTM/GRU initialized parameters. Results are the average of the 10 studied instances and the shaded region shows the standard error. Bar plots in the right corner show the results of last step of optimization. Adagrad optimizer with learning rate of 0.1 is used as a classical optimization routine for both QAOA and DC-QAOA. We observe that both the methods successfully find the initial parameters, and require a few iterations to get to the minima by the classical optimizer.
	
	For results of 3-regular MaxCut shown in Fig. ~\ref{fig:MaxCut} (a) we can see that LSTM initialized DC-QAOA works better in the sense that can reach to lower $\mathcal{E}$ values with fewer iterations. As far as the GRU is concerned, GRU with QAOA works better for this problem. Hence, as expected we can achieve lower $\mathcal{E}$ with only a few iterations by applying this technique.
	Regarding the complete weighted case results shown in Fig.~\ref{fig:MaxCut} (b), a recent study on the optimization landscape for QAOA for weighted and unweighted MaxCut ~\cite{shaydulin2022parameter} stating that weighted edges lead to parameter landscapes with lots of local optima. Due to this reason absolute $\mathcal{E}$ values are higher as compared to the unweighted case. We can see from the bar plots showing error after 200 iterations LSTM and GRU intialized DC-QAOA are the best methods. This also shows that GRU can achieve a similar performance as LSTM with a fewer parameters so GRU has more scalability with respect to the system size. Again, RNN initialization outperforms random initialization in terms of final result and convergence. It is also remarkable that if we look at the RNN cells as black-box optimizers, they reach values very close to the convergence with a small horizon time ($T=6$).  As an additional analysis, we also study the parameter concentration phenomenon in DC-QAOA. Parameter concentration is a effect where for some class of problems, if the parameter values are fixed, different instances will have nearly the same objective function values~\cite{brandao2018fixed}. In other words, the optimal parameter values for these instances will be similar. Parameter concentration has already been analyzed for QAOA with respect to MaxCut~\cite{brandao2018fixed} and SK model~\cite{PhysRevA.104.L010401} . Meta-Learning of RNNs also shows this behaviour~\cite{verdon2019learning}. Recently, it was observed that this effect aids in alleviating barren plateaus~\cite{mele2022avoiding}. These concentrations can serve as a good initialization strategy as the optimal parameters of small system sizes can be utilized as initial parameters for larger complex systems. We examine the parameter concentration effect in DC-QAOA for unweighted 3-regular MaxCut problems to check if this effect is still intact after adding the CD-term and if it gives any advantage over QAOA.
	
	\begin{figure}
		\centering
		\includegraphics[width = 1\linewidth]{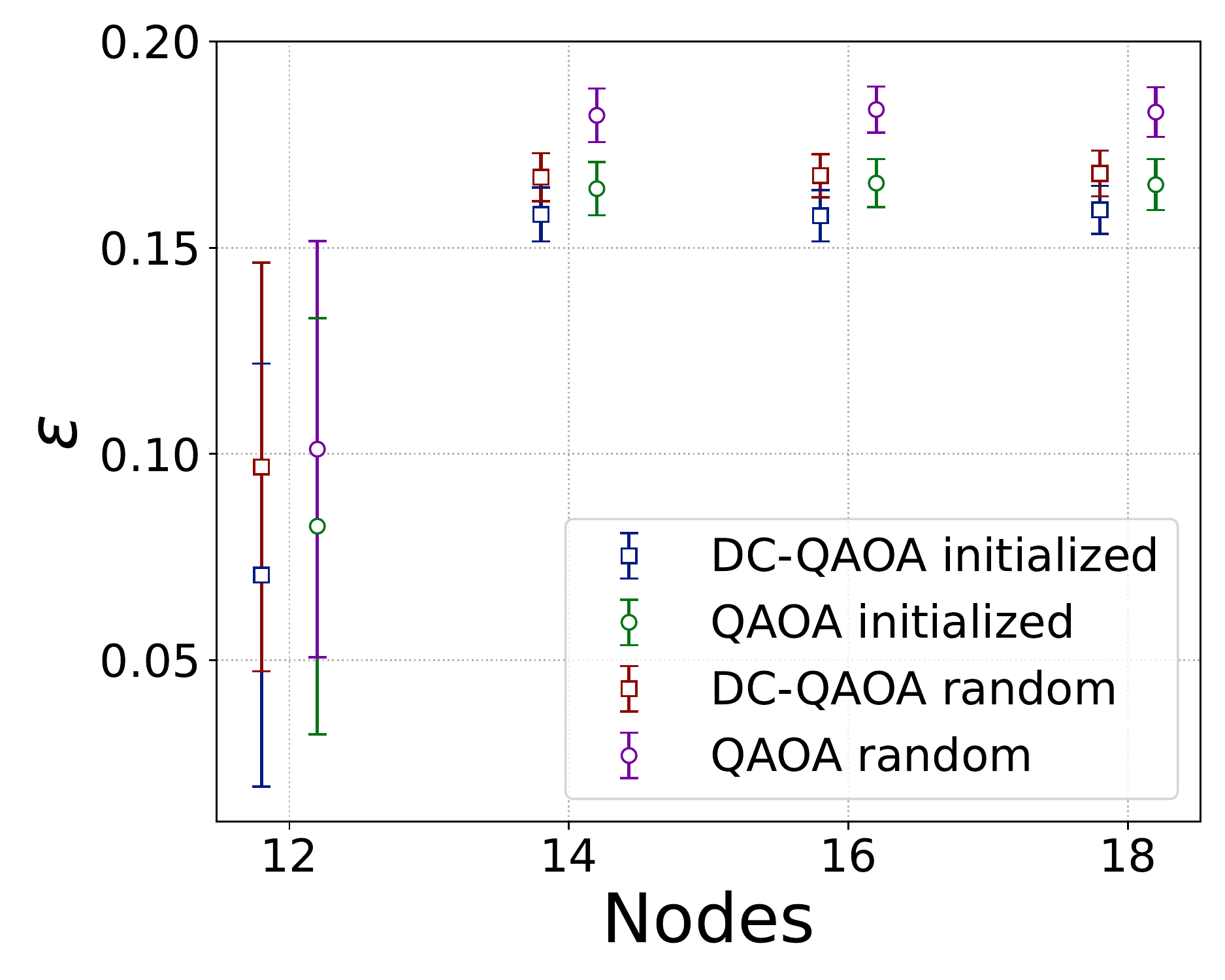}
		\caption{Relative error ($\mathcal{E}$) as a function of the nodes for 3-regular unweighted MaxCut problem with $p=2$. DC-QAOA (QAOA) `initialized' shows $\mathcal{E}$ values where the optimal parameters obtained with LSTM for a 10-qubit system were used as the initial parameters for higher system DC-QAOA (QAOA) whereas `random' show random initialization. Results are the average over 10 random instances of graphs and 100 iteration steps with error bars showing standard error.}
		\label{paramsconc}
	\end{figure}
	
	In Fig.~\ref{paramsconc}, we plot $\mathcal{E}$ as a function of the nodes of the graph for $p=2$ 3-regular unweighted MaxCut problems. This plot compares 10-qubit initialized graphs where the optimal parameters for 10 node system obtained with LSTM were used as initial parameters for 14, 16, and 18 node systems with randomly initialized graphs where parameters were initialized randomly. $\mathcal{E}$ shows mean of 10 graph instances with 100 steps of iteration. For both QAOA and DC-QAOA, we can observe the parameter concentration effect, i.e. $\mathcal{E}$ values for the 10qubit-initialized graphs have relatively lower $\mathcal{E}$ values as compared to randomly initialized graphs. As expected, DC-QAOA $\mathcal{E}$ values are the lowest implying that concentrations strategy combined with DC-QAOA have advantages over QAOA. We can also show that as the system size increase $\mathcal{E}$ values start to saturate. This can be attributed to the fact that for higher system sizes, the expressibility for QAOA and DC-QAOA decrease significantly and thus the ansatz might not be expressive to explore larger Hilbert space. This performance might improve if instead of using optimal parameters from 10 node system, 12 or 14 nodes system is considered. Nevertheless, in conclusion, the parameter concentration effect is also present in DC-QAOA and serves as a good initialization strategy in the sense that LSTM or GRU can be implemented for a smaller system and the optimal parameters can be used for larger systems.

	\subsection{The SK model}
	Second benchmark for our technique is the SK model, which is a classical spin-model with couplings that possess an all-to-all connectivity~\cite{RevModPhys.58.801,PhysRevLett.35.1792}. This model is important to investigate because lots of optimization problems can be encoded into spin glass systems~\cite{10.3389/fphy.2014.00005}. The Hamiltonian of the SK model is given by,
	\begin{equation}
		H_c=-\sum_{i, j } J_{i j} \sigma_{z}^{i} \sigma_{z}^{j}
	\end{equation}
	where $J_{ij}$ are coupling coefficients associated to spin $i$ and spin $j$. We study a 10 qubit SK model with $J_{ij} \in \{-1,1\}$ with mean 0 and variance 1. There has been substantial research applying QAOA to SK model recently~\cite{Harrigan2021,farhi2019quantum2}. The CD unitary for that works best for this case will be again $U_D(\alpha) = e^{-i\alpha \sum_{ij} J_{ij} \sigma_z^i\sigma_y^j}$ (Appendix ~\ref{appendixB}). As far as the training of LSTM is concerned, we selected 200 graphs with all-to-all connectivity and the coupling coefficients $J_{ij}$ generated randomly from $J_{ij} \in \{-1,1\}$. Similar to MaxCut problem, the maximum epochs were kept to 10 with $\tau=0.01$, the time horizon was set to $T=6$, the loss function is given by Eq.~\eqref{lossfn} and Adam optimizer with step size=0.01 was used to train the network. To do the final optimization, Adagrad optimizer with 0.1 learning rate was used.
	\begin{figure}[t]
		\centering
		\includegraphics[width=1\linewidth]{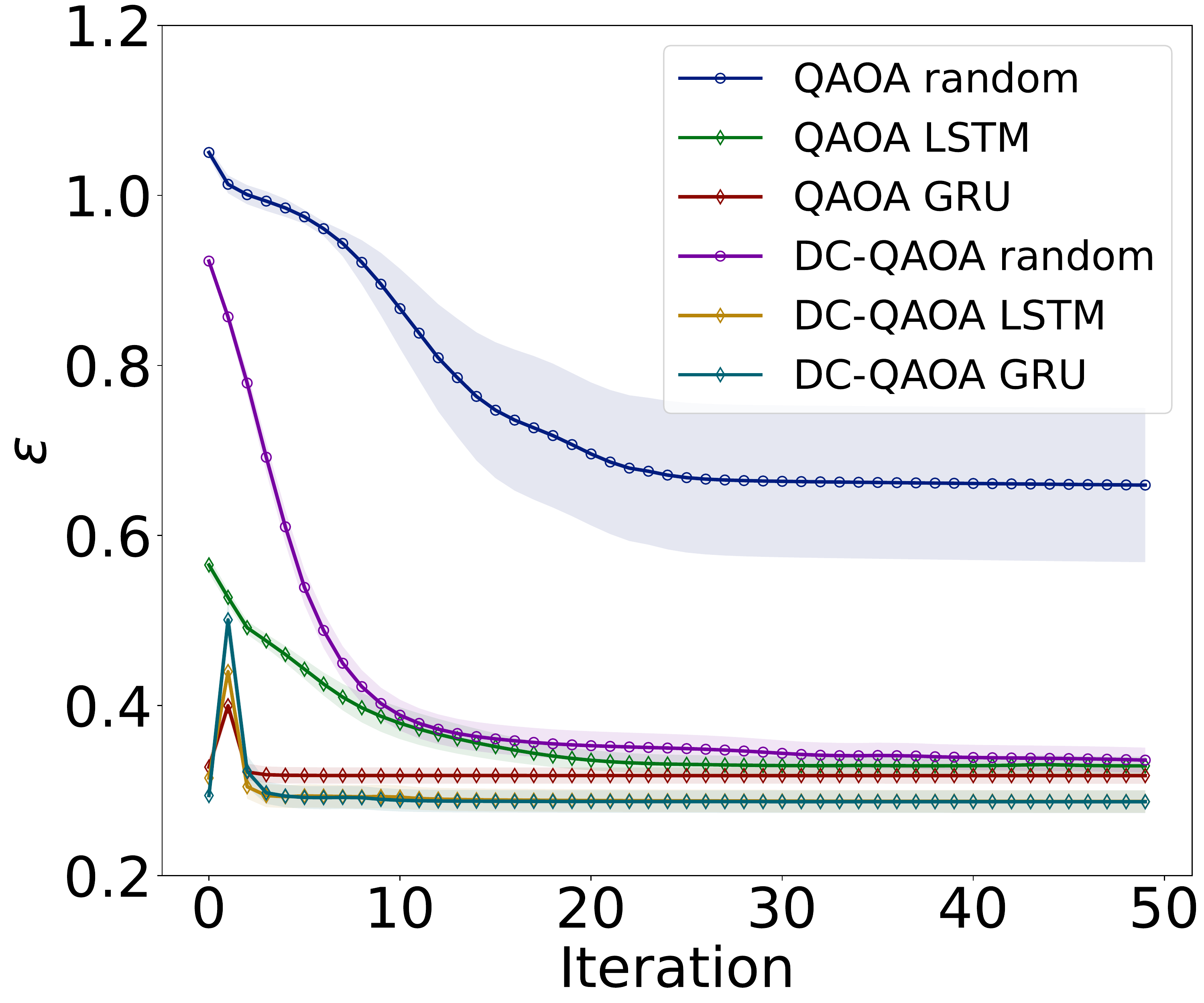}
		\caption{Relative error ($\mathcal{E}$) as a function of number of iterations for 10 instances of the SK model. Comparison of QAOA and DC-QAOA  for $p=2$ with random seeds and LSTM/GRU initialized seeds shown by different colored lines. Shaded area shows standard error. Results were obtained on an ideal quantum simulator.}
		\label{fig:SKres}
	\end{figure}
	
	In Fig.~\ref{fig:SKres}, we illustrate $\mathcal{E}$ of 10 random graph instances with respect to the number of steps for different initializations of QAOA and DC-QAOA  for $p=2$. LSTM and GRU initialization of optimizations for DC-QAOA problems shows best results in terms of convergence speed and final relative error. One remarkable thing is that as in the MaxCut case, GRU-QAOA has a faster convergence than LSTM-QAOA, which may suggest a better performance of the first one in the QAOA case. Once again, DC-QAOA has outperformed QAOA for the same initialization. Similarly to  MaxCut, the optimization of the network itself reaches values quite close to the final relative error.

	\section{Discussion and Conclusion} \label{disc}
	We have proposed a meta-learning technique as an initialization strategy for DC-QAOA. Particularly,  RNNs, i.e., LSTMs and GRUs are employed to design a black-box classical optimization routine that learns from gradients to optimize a loss function and outputs optimal parameters that can be used as initial parameters for DC-QAOA. To benchmark this, we have exemplified unweighted 3-regular MaxCut, weighted complete MaxCut and the SK model. For all the models we study, RNN initialization combined with DC-QAOA outperforms QAOA in the sense that we can reach to lower relative errors ($\mathcal{E}$) in lower iterations. We also investigate the parameter concentration effect on DC-QAOA and show that this can  pose as a good alternative initialization strategy. The two major challenges of the near-term era i.e. getting higher success and finding optimal tunable parameters to reduce the optimization overload are addressed here. We also show that GRUs can perform as well as LSTMs with a lower number of cell operations which suggests that GRUs might be a better choice of RNNs for this kind of meta-learning techniques.
	
	For future work, application of this method for industrial problems like portfolio optimization would be interesting. Also, the performance with higher layers and higher qubit systems may be studied. In conclusion, digitized-counterdiabatic quantum computing (DCQC) is a new paradigm that has already reported many studies showing improvement over current adiabatic optimization methods~\cite{chandarana2022digitized,hegadePRAletter,hegade2021portfolio,hegade2021shortcuts,hegade2022digitized}, still there is a lot of room for improving these algorithms and implementing sophisticated machine learning methods might as well be a key reaching nearer to quantum advantage in the current noisy intermediate-scale quantum  era. Also, this method has a possible implication in real hardware implementation, because the presence of several noise sources that may change the performance of the RNNs.

	\begin{acknowledgments}
		
		This work is supported by EU FET Open Grant EPIQUS (899368) and Quromorphic (828826), QUANTEK project (KK-2021/00070) and NSFC (12075145 and 12211540002).
		X.C. acknowledges the Ram\'on y Cajal program (RYC-2017-22482).
		
	\end{acknowledgments}
	
	\appendix
	\section{Comparison of different CD terms} \label{appendixB}
	
	\begin{figure}[ht]
		\centering
		\includegraphics[width=0.48\linewidth]{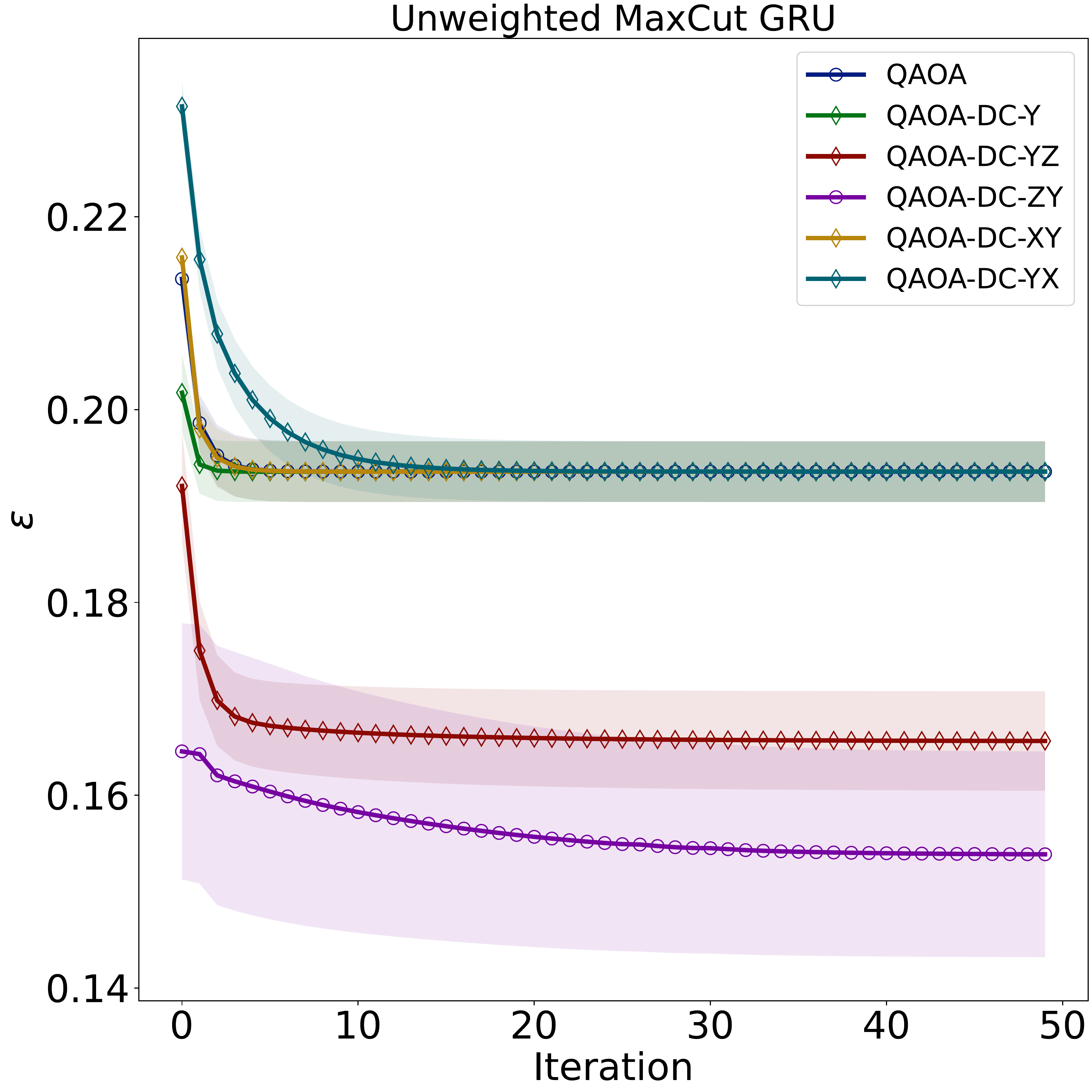}
		\includegraphics[width=0.48\linewidth]{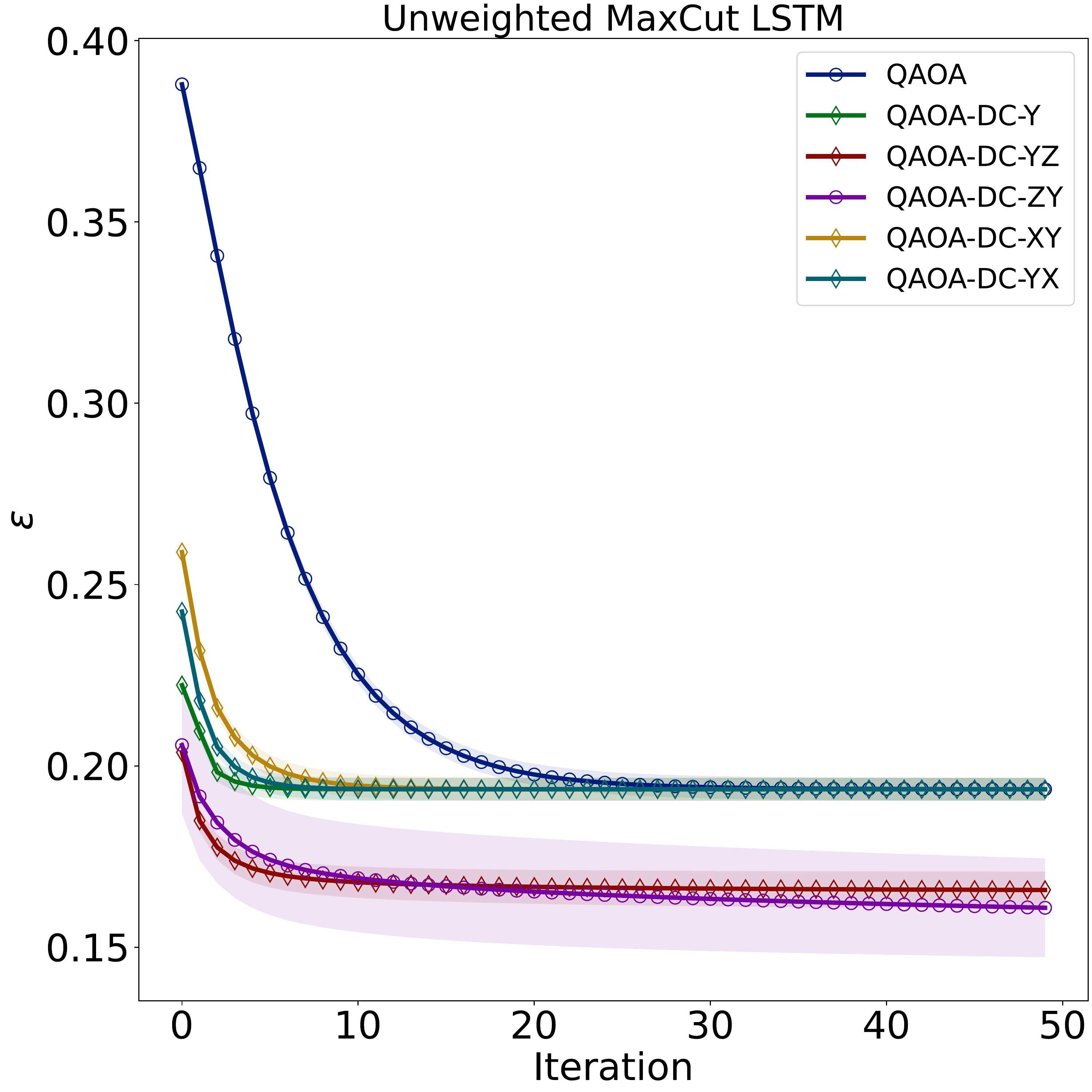}
		\includegraphics[width=0.48\linewidth]{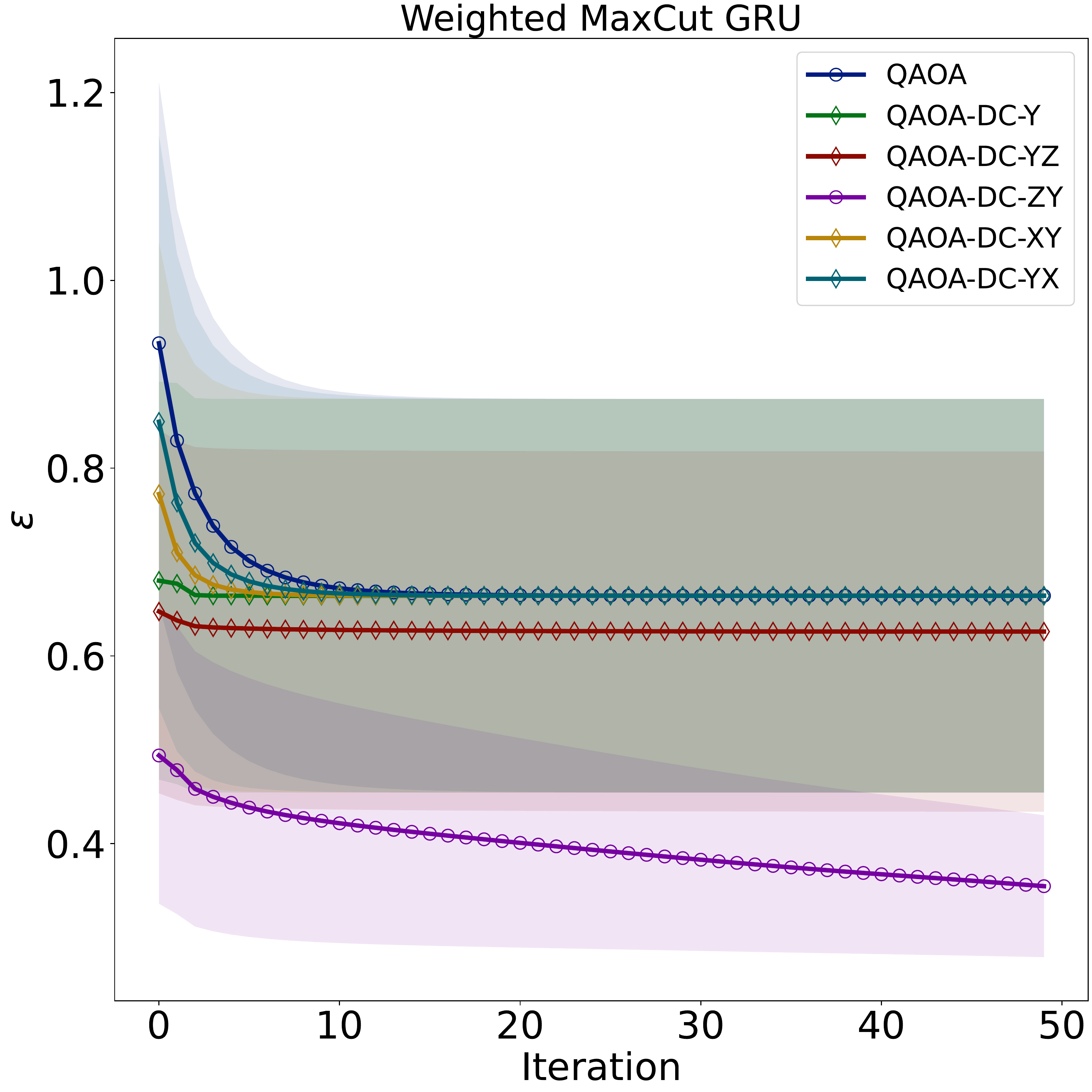}
		\includegraphics[width=0.48\linewidth]{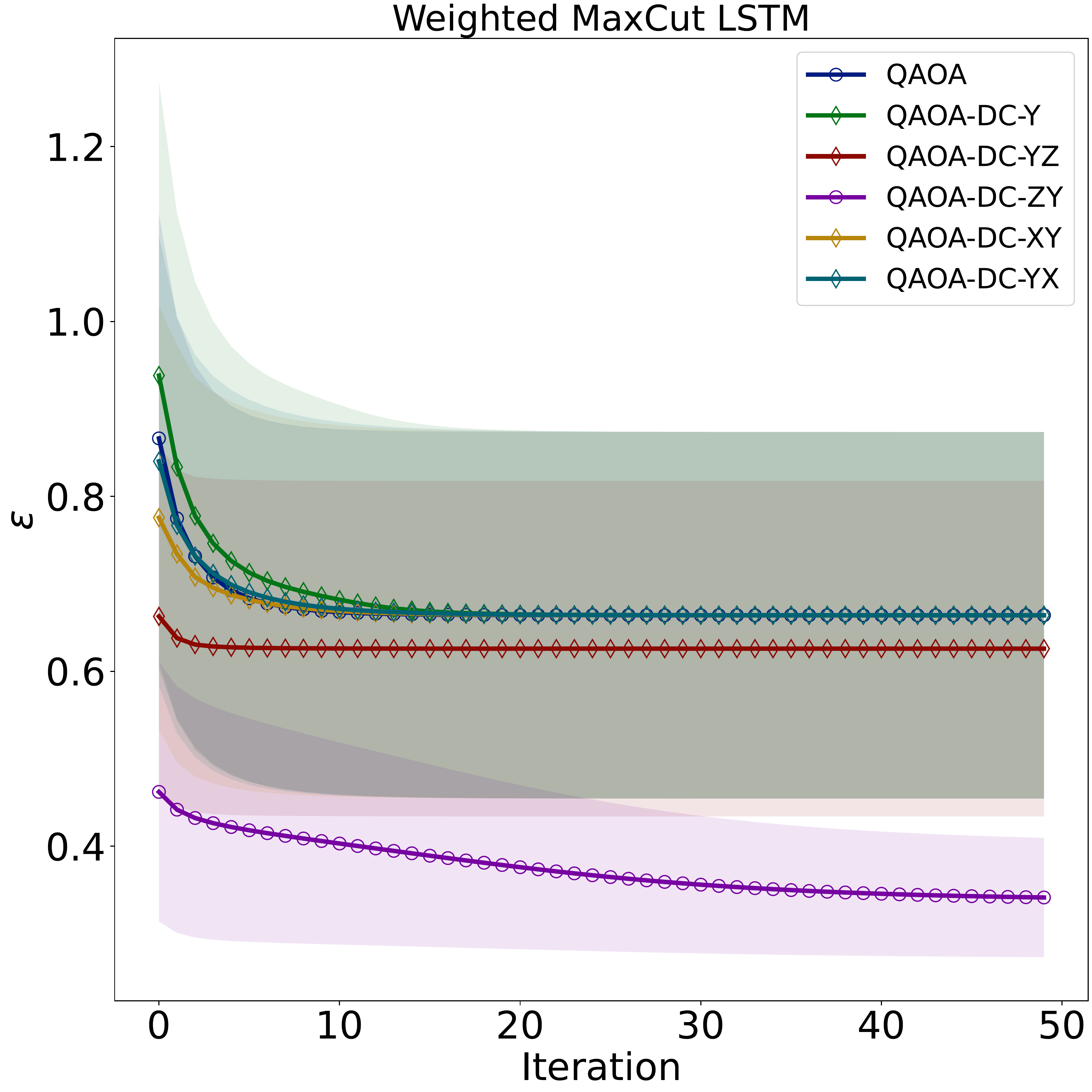}
		\includegraphics[width=0.48\linewidth]{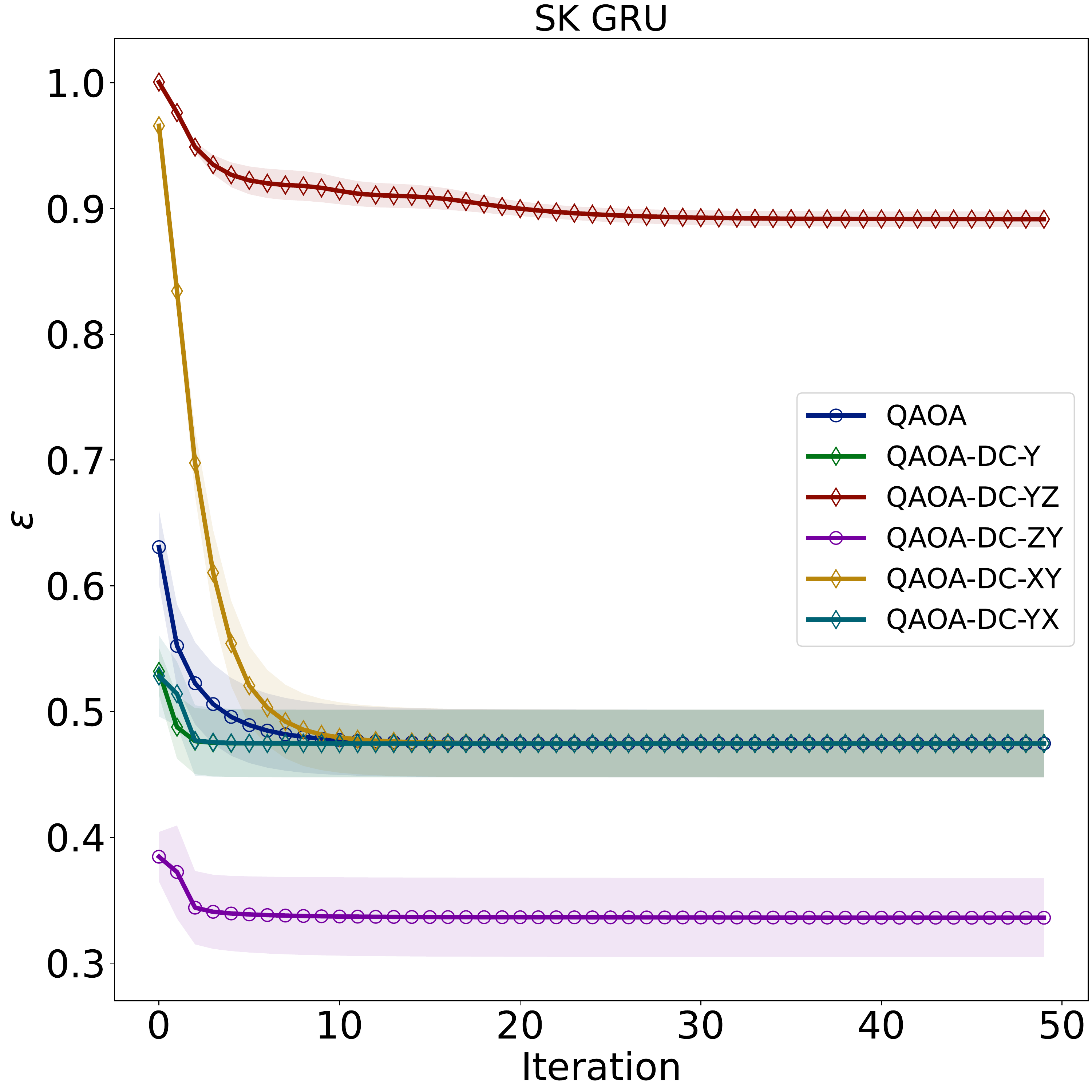}
		\includegraphics[width=0.48\linewidth]{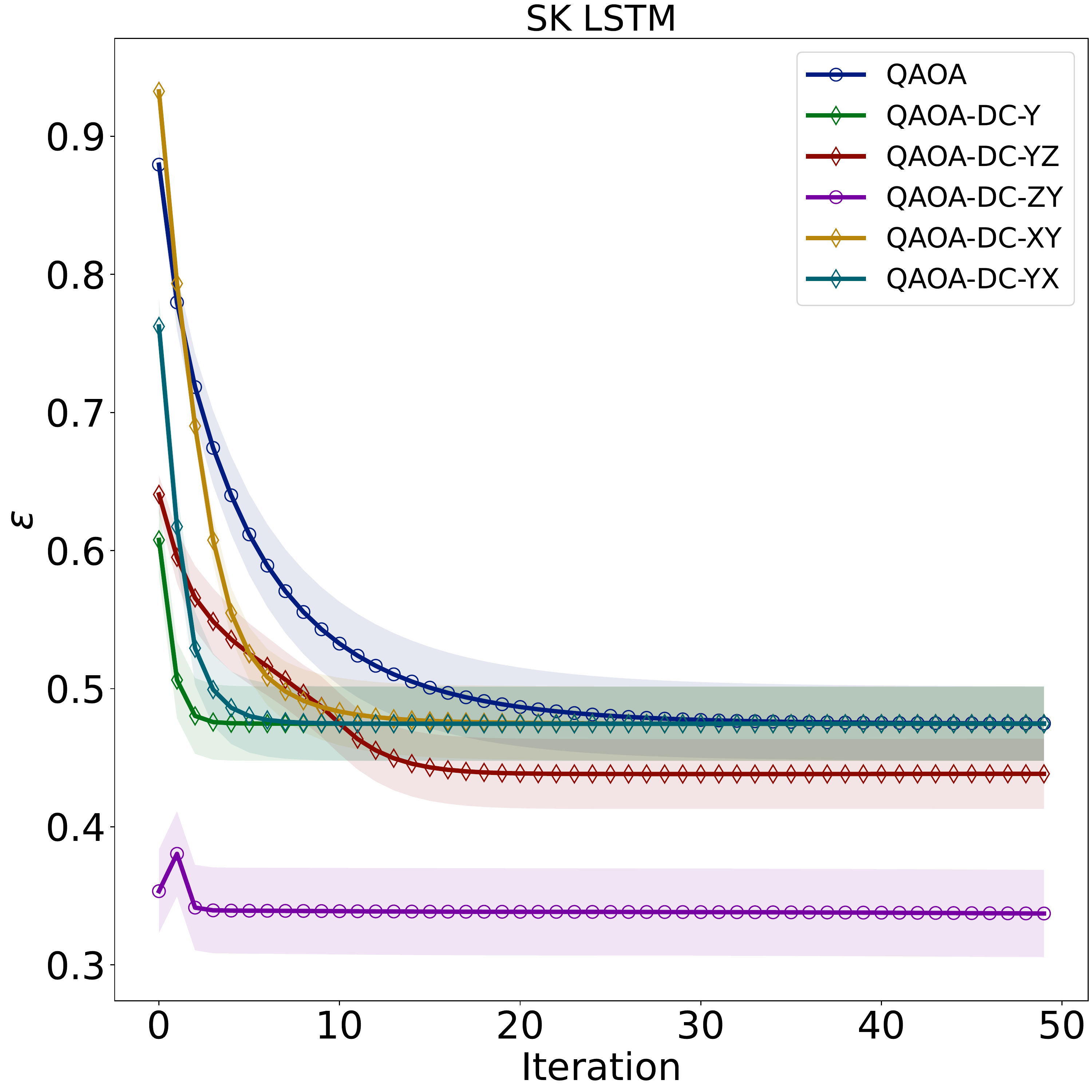}
		\label{fig:CD_comparison}
		\caption{Comparison of performance of different CD terms. Mean $\mathcal{E}$ for 5 graphs vs. number of iterations is plotted for 8 qubit system and $p=1$ layer. A train sample of 100 graphs was used for training the cells. }
	\end{figure}
	
	The CD terms used in this study was obtained from the NC method (Eq.~\ref{gauge}). In Fig.~\ref{fig:CD_comparison}, we have chosen a 8 qubit system with $p=1$, with an aim to compare all the second order CD terms obtained from Eq.~\ref{gauge}. We can see that $A^{(2)}_{\lambda} = \{ \sigma^z \sigma^y\}$ performs exceptionally well for all the models and thus we have implemented the same for our numerical analysis for higher qubit systems.
	
	\section{Information flow in LSTM and GRU}\label{Appinfo}
	The standard equations that describe information flow one layer of LSTM network are the following :  
	\begin{subequations}
		\begin{align}
			f_t \; =& \; \sigma\left(W_f \theta_t + R_fh_{t-1} + b_f\right), \label{eq:LSTM_1} \\
			i_t \; =& \; \sigma\left(W_i \theta_t + R_ih_{t-1} + b_i\right), \label{eq:LSTM_2} \\ 
			\widetilde{c}_t \; =& \; tanh\left(W_c \theta_t + R_ch_{t-1}+ b_c \right) , \label{eq:LSTM_3} \\
			c_t \; =& \; f_t \odot c_{t-1} +i_t \odot \widetilde{c}_t , \label{eq:LSTM_4} \\ 
			o_t \; =& \; \sigma\left(W_o \theta_t + R_oh_{t-1} + b_o\right), \label{eq:LSTM_5} \\
			h_t \; =& \; o_t \odot tanh\left(c_t\right), \label{eq:LSTM_6}
		\end{align}
		\label{eq:LSTM}
	\end{subequations}
	where $\theta_t$ is the input vector on each time step  and $h_{t-1}$ the previous hidden state. $W$ and $R$ are weights to the input and hidden state and $b$ are the biases vectors. $W$, $R$ and $b$ are learned during the training of the network by minimizing some loss function $\mathcal{L}({\hat{y}})$. $\sigma$ and $
	\tanh$ are the sigmoid and hyperbolic tangent functions. As their ranges are $[0,1]$ and $[-1,1]$ they are useful to keep or discard information and regulate the network. $\odot$ denotes the Hadamard product.
	Eq.~\eqref{eq:LSTM} shows how the information flows throughout the network. Forget gate $f_t$ decides how much information from the previous cell state $c_t$ is kept at time $t$. Similarly input gate $i_t$ multiplies the \textit{update gate} $\widetilde{c}_t $ in order to regulate the information given by $\theta_t$ and $h_{t-1}$. Finally, the output gate $o_t$ will decide the information kept in the hidden state for the next time step $h_t$.
	For the case of GRU cell, the equations have different forms from which we implement the following: 
	\begin{subequations}
		\begin{align}
			z_t \; =& \; \sigma\left(W_z \theta_t + R_zh_{t-1} + b_z + d_z\right), \label{eq:GRU_1} \\
			r_t \; =& \; \sigma\left( W_r \theta_t + R_rh_{t-1} + b_r + d_r\right), \label{eq:GRU_2} \\ 
			\widetilde{h}_t \; =& \; tanh\left(W_h \theta_t + r_t\odot (R_hh_{t-1}+d_h) +b_h \right), \label{eq:GRU_3} \\ 
			h_t \; =& \; z_t \odot h_{t-1} + \left(1-z_t\right) \odot \widetilde{h}_{t}. \label{eq:GRU_4}
		\end{align}
		\label{eq:GRU}
	\end{subequations}
	As it can be seen the gates will be different now. First of all, another bias vector $d$ is introduced that will help to modulate the calculation of $\widetilde{h}_t$. Update gate $z_t$ will select information from previous hidden state. Similarly, the reset gate $r_t$ will select which information to discard from previous hidden states. $\widetilde{h}_t$ is known as candidate state gate and it's used to generate a candidate of a new hidden state. It can be seen how the reset gate helps to forget unimportant information from previous iterations. Finally, to compute the new hidden state $h_t$, the update gate will be considered, and will model how much weight is given to the candidate state $\widetilde{h}_t$ and to the previous hidden state $h_{t-1}$.

	\bibliography{main.bib}

\begin{thebibliography}{86}%
\makeatletter
\providecommand \@ifxundefined [1]{%
 \@ifx{#1\undefined}
}%
\providecommand \@ifnum [1]{%
 \ifnum #1\expandafter \@firstoftwo
 \else \expandafter \@secondoftwo
 \fi
}%
\providecommand \@ifx [1]{%
 \ifx #1\expandafter \@firstoftwo
 \else \expandafter \@secondoftwo
 \fi
}%
\providecommand \natexlab [1]{#1}%
\providecommand \enquote  [1]{``#1''}%
\providecommand \bibnamefont  [1]{#1}%
\providecommand \bibfnamefont [1]{#1}%
\providecommand \citenamefont [1]{#1}%
\providecommand \href@noop [0]{\@secondoftwo}%
\providecommand \href [0]{\begingroup \@sanitize@url \@href}%
\providecommand \@href[1]{\@@startlink{#1}\@@href}%
\providecommand \@@href[1]{\endgroup#1\@@endlink}%
\providecommand \@sanitize@url [0]{\catcode `\\12\catcode `\$12\catcode
  `\&12\catcode `\#12\catcode `\^12\catcode `\_12\catcode `\%12\relax}%
\providecommand \@@startlink[1]{}%
\providecommand \@@endlink[0]{}%
\providecommand \url  [0]{\begingroup\@sanitize@url \@url }%
\providecommand \@url [1]{\endgroup\@href {#1}{\urlprefix }}%
\providecommand \urlprefix  [0]{URL }%
\providecommand \Eprint [0]{\href }%
\providecommand \doibase [0]{https://doi.org/}%
\providecommand \selectlanguage [0]{\@gobble}%
\providecommand \bibinfo  [0]{\@secondoftwo}%
\providecommand \bibfield  [0]{\@secondoftwo}%
\providecommand \translation [1]{[#1]}%
\providecommand \BibitemOpen [0]{}%
\providecommand \bibitemStop [0]{}%
\providecommand \bibitemNoStop [0]{.\EOS\space}%
\providecommand \EOS [0]{\spacefactor3000\relax}%
\providecommand \BibitemShut  [1]{\csname bibitem#1\endcsname}%
\let\auto@bib@innerbib\@empty
\bibitem [{\citenamefont {Peruzzo}\ \emph {et~al.}(2014)\citenamefont
  {Peruzzo}, \citenamefont {McClean}, \citenamefont {Shadbolt}, \citenamefont
  {Yung}, \citenamefont {Zhou}, \citenamefont {Love}, \citenamefont
  {Aspuru-Guzik},\ and\ \citenamefont {O’brien}}]{peruzzo2014variational}%
  \BibitemOpen
  \bibfield  {author} {\bibinfo {author} {\bibfnamefont {A.}~\bibnamefont
  {Peruzzo}}, \bibinfo {author} {\bibfnamefont {J.}~\bibnamefont {McClean}},
  \bibinfo {author} {\bibfnamefont {P.}~\bibnamefont {Shadbolt}}, \bibinfo
  {author} {\bibfnamefont {M.-H.}\ \bibnamefont {Yung}}, \bibinfo {author}
  {\bibfnamefont {X.-Q.}\ \bibnamefont {Zhou}}, \bibinfo {author}
  {\bibfnamefont {P.~J.}\ \bibnamefont {Love}}, \bibinfo {author}
  {\bibfnamefont {A.}~\bibnamefont {Aspuru-Guzik}},\ and\ \bibinfo {author}
  {\bibfnamefont {J.~L.}\ \bibnamefont {O’brien}},\ }\bibfield  {title}
  {\bibinfo {title} {A variational eigenvalue solver on a photonic quantum
  processor},\ }\href {https://doi.org/10.1038/ncomms5213} {\bibfield
  {journal} {\bibinfo  {journal} {Nature communications}\ }\textbf {\bibinfo
  {volume} {5}},\ \bibinfo {pages} {1} (\bibinfo {year} {2014})}\BibitemShut
  {NoStop}%
\bibitem [{\citenamefont {Killoran}\ \emph {et~al.}(2019)\citenamefont
  {Killoran}, \citenamefont {Bromley}, \citenamefont {Arrazola}, \citenamefont
  {Schuld}, \citenamefont {Quesada},\ and\ \citenamefont
  {Lloyd}}]{killoran2019continuous}%
  \BibitemOpen
  \bibfield  {author} {\bibinfo {author} {\bibfnamefont {N.}~\bibnamefont
  {Killoran}}, \bibinfo {author} {\bibfnamefont {T.~R.}\ \bibnamefont
  {Bromley}}, \bibinfo {author} {\bibfnamefont {J.~M.}\ \bibnamefont
  {Arrazola}}, \bibinfo {author} {\bibfnamefont {M.}~\bibnamefont {Schuld}},
  \bibinfo {author} {\bibfnamefont {N.}~\bibnamefont {Quesada}},\ and\ \bibinfo
  {author} {\bibfnamefont {S.}~\bibnamefont {Lloyd}},\ }\bibfield  {title}
  {\bibinfo {title} {Continuous-variable quantum neural networks},\ }\href
  {https://doi.org/10.1103/PhysRevResearch.1.033063} {\bibfield  {journal}
  {\bibinfo  {journal} {Physical Review Research}\ }\textbf {\bibinfo {volume}
  {1}},\ \bibinfo {pages} {033063} (\bibinfo {year} {2019})}\BibitemShut
  {NoStop}%
\bibitem [{\citenamefont {Wecker}\ \emph {et~al.}(2015)\citenamefont {Wecker},
  \citenamefont {Hastings},\ and\ \citenamefont {Troyer}}]{wecker2015progress}%
  \BibitemOpen
  \bibfield  {author} {\bibinfo {author} {\bibfnamefont {D.}~\bibnamefont
  {Wecker}}, \bibinfo {author} {\bibfnamefont {M.~B.}\ \bibnamefont
  {Hastings}},\ and\ \bibinfo {author} {\bibfnamefont {M.}~\bibnamefont
  {Troyer}},\ }\bibfield  {title} {\bibinfo {title} {Progress towards practical
  quantum variational algorithms},\ }\href
  {https://doi.org/10.1103/PhysRevA.92.042303} {\bibfield  {journal} {\bibinfo
  {journal} {Physical Review A}\ }\textbf {\bibinfo {volume} {92}},\ \bibinfo
  {pages} {042303} (\bibinfo {year} {2015})}\BibitemShut {NoStop}%
\bibitem [{\citenamefont {Biamonte}\ \emph {et~al.}(2017)\citenamefont
  {Biamonte}, \citenamefont {Wittek}, \citenamefont {Pancotti}, \citenamefont
  {Rebentrost}, \citenamefont {Wiebe},\ and\ \citenamefont
  {Lloyd}}]{biamonte2017quantum}%
  \BibitemOpen
  \bibfield  {author} {\bibinfo {author} {\bibfnamefont {J.}~\bibnamefont
  {Biamonte}}, \bibinfo {author} {\bibfnamefont {P.}~\bibnamefont {Wittek}},
  \bibinfo {author} {\bibfnamefont {N.}~\bibnamefont {Pancotti}}, \bibinfo
  {author} {\bibfnamefont {P.}~\bibnamefont {Rebentrost}}, \bibinfo {author}
  {\bibfnamefont {N.}~\bibnamefont {Wiebe}},\ and\ \bibinfo {author}
  {\bibfnamefont {S.}~\bibnamefont {Lloyd}},\ }\bibfield  {title} {\bibinfo
  {title} {Quantum machine learning},\ }\href
  {https://doi.org/10.1038/nature23474} {\bibfield  {journal} {\bibinfo
  {journal} {Nature}\ }\textbf {\bibinfo {volume} {549}},\ \bibinfo {pages}
  {195} (\bibinfo {year} {2017})}\BibitemShut {NoStop}%
\bibitem [{\citenamefont {Zhou}\ \emph
  {et~al.}(2020{\natexlab{a}})\citenamefont {Zhou}, \citenamefont {Wang},
  \citenamefont {Choi}, \citenamefont {Pichler},\ and\ \citenamefont
  {Lukin}}]{zhou2020quantum}%
  \BibitemOpen
  \bibfield  {author} {\bibinfo {author} {\bibfnamefont {L.}~\bibnamefont
  {Zhou}}, \bibinfo {author} {\bibfnamefont {S.-T.}\ \bibnamefont {Wang}},
  \bibinfo {author} {\bibfnamefont {S.}~\bibnamefont {Choi}}, \bibinfo {author}
  {\bibfnamefont {H.}~\bibnamefont {Pichler}},\ and\ \bibinfo {author}
  {\bibfnamefont {M.~D.}\ \bibnamefont {Lukin}},\ }\bibfield  {title} {\bibinfo
  {title} {Quantum approximate optimization algorithm: Performance, mechanism,
  and implementation on near-term devices},\ }\href
  {https://doi.org/10.1103/PhysRevX.10.021067} {\bibfield  {journal} {\bibinfo
  {journal} {Physical Review X}\ }\textbf {\bibinfo {volume} {10}},\ \bibinfo
  {pages} {021067} (\bibinfo {year} {2020}{\natexlab{a}})}\BibitemShut
  {NoStop}%
\bibitem [{\citenamefont {Bharti}\ \emph {et~al.}(2022)\citenamefont {Bharti},
  \citenamefont {Cervera-Lierta}, \citenamefont {Kyaw}, \citenamefont {Haug},
  \citenamefont {Alperin-Lea}, \citenamefont {Anand}, \citenamefont {Degroote},
  \citenamefont {Heimonen}, \citenamefont {Kottmann}, \citenamefont {Menke},
  \citenamefont {Mok}, \citenamefont {Sim}, \citenamefont {Kwek},\ and\
  \citenamefont {Aspuru-Guzik}}]{RevModPhys.94.015004}%
  \BibitemOpen
  \bibfield  {author} {\bibinfo {author} {\bibfnamefont {K.}~\bibnamefont
  {Bharti}}, \bibinfo {author} {\bibfnamefont {A.}~\bibnamefont
  {Cervera-Lierta}}, \bibinfo {author} {\bibfnamefont {T.~H.}\ \bibnamefont
  {Kyaw}}, \bibinfo {author} {\bibfnamefont {T.}~\bibnamefont {Haug}}, \bibinfo
  {author} {\bibfnamefont {S.}~\bibnamefont {Alperin-Lea}}, \bibinfo {author}
  {\bibfnamefont {A.}~\bibnamefont {Anand}}, \bibinfo {author} {\bibfnamefont
  {M.}~\bibnamefont {Degroote}}, \bibinfo {author} {\bibfnamefont
  {H.}~\bibnamefont {Heimonen}}, \bibinfo {author} {\bibfnamefont {J.~S.}\
  \bibnamefont {Kottmann}}, \bibinfo {author} {\bibfnamefont {T.}~\bibnamefont
  {Menke}}, \bibinfo {author} {\bibfnamefont {W.-K.}\ \bibnamefont {Mok}},
  \bibinfo {author} {\bibfnamefont {S.}~\bibnamefont {Sim}}, \bibinfo {author}
  {\bibfnamefont {L.-C.}\ \bibnamefont {Kwek}},\ and\ \bibinfo {author}
  {\bibfnamefont {A.}~\bibnamefont {Aspuru-Guzik}},\ }\bibfield  {title}
  {\bibinfo {title} {Noisy intermediate-scale quantum algorithms},\ }\href
  {https://doi.org/10.1103/RevModPhys.94.015004} {\bibfield  {journal}
  {\bibinfo  {journal} {Reviews of Modern Physics}\ }\textbf {\bibinfo {volume}
  {94}},\ \bibinfo {pages} {015004} (\bibinfo {year} {2022})}\BibitemShut
  {NoStop}%
\bibitem [{\citenamefont {Farhi}\ \emph {et~al.}(2014)\citenamefont {Farhi},
  \citenamefont {Goldstone},\ and\ \citenamefont {Gutmann}}]{farhi2014quantum}%
  \BibitemOpen
  \bibfield  {author} {\bibinfo {author} {\bibfnamefont {E.}~\bibnamefont
  {Farhi}}, \bibinfo {author} {\bibfnamefont {J.}~\bibnamefont {Goldstone}},\
  and\ \bibinfo {author} {\bibfnamefont {S.}~\bibnamefont {Gutmann}},\
  }\bibfield  {title} {\bibinfo {title} {A quantum approximate optimization
  algorithm},\ }\href {https://arxiv.org/abs/1411.4028} {\bibfield  {journal}
  {\bibinfo  {journal} {arXiv preprint arXiv:1411.4028}\ } (\bibinfo {year}
  {2014})}\BibitemShut {NoStop}%
\bibitem [{\citenamefont {Farhi}\ and\ \citenamefont
  {Harrow}(2016)}]{farhi2016quantum}%
  \BibitemOpen
  \bibfield  {author} {\bibinfo {author} {\bibfnamefont {E.}~\bibnamefont
  {Farhi}}\ and\ \bibinfo {author} {\bibfnamefont {A.~W.}\ \bibnamefont
  {Harrow}},\ }\bibfield  {title} {\bibinfo {title} {Quantum supremacy through
  the quantum approximate optimization algorithm},\ }\href
  {https://arxiv.org/abs/1602.07674} {\bibfield  {journal} {\bibinfo  {journal}
  {arXiv preprint arXiv:1602.07674}\ } (\bibinfo {year} {2016})}\BibitemShut
  {NoStop}%
\bibitem [{\citenamefont {Gu{\'e}ry-Odelin}\ \emph {et~al.}(2019)\citenamefont
  {Gu{\'e}ry-Odelin}, \citenamefont {Ruschhaupt}, \citenamefont {Kiely},
  \citenamefont {Torrontegui}, \citenamefont {Mart{\'\i}nez-Garaot},\ and\
  \citenamefont {Muga}}]{guery2019shortcuts}%
  \BibitemOpen
  \bibfield  {author} {\bibinfo {author} {\bibfnamefont {D.}~\bibnamefont
  {Gu{\'e}ry-Odelin}}, \bibinfo {author} {\bibfnamefont {A.}~\bibnamefont
  {Ruschhaupt}}, \bibinfo {author} {\bibfnamefont {A.}~\bibnamefont {Kiely}},
  \bibinfo {author} {\bibfnamefont {E.}~\bibnamefont {Torrontegui}}, \bibinfo
  {author} {\bibfnamefont {S.}~\bibnamefont {Mart{\'\i}nez-Garaot}},\ and\
  \bibinfo {author} {\bibfnamefont {J.~G.}\ \bibnamefont {Muga}},\ }\bibfield
  {title} {\bibinfo {title} {Shortcuts to adiabaticity: Concepts, methods, and
  applications},\ }\href {https://doi.org/10.1103/RevModPhys.91.045001}
  {\bibfield  {journal} {\bibinfo  {journal} {Reviews of Modern Physics}\
  }\textbf {\bibinfo {volume} {91}},\ \bibinfo {pages} {045001} (\bibinfo
  {year} {2019})}\BibitemShut {NoStop}%
\bibitem [{\citenamefont {Torrontegui}\ \emph {et~al.}(2013)\citenamefont
  {Torrontegui}, \citenamefont {Ib{\'a}nez}, \citenamefont
  {Mart{\'\i}nez-Garaot}, \citenamefont {Modugno}, \citenamefont {del Campo},
  \citenamefont {Gu{\'e}ry-Odelin}, \citenamefont {Ruschhaupt}, \citenamefont
  {Chen},\ and\ \citenamefont {Muga}}]{torrontegui2013shortcuts}%
  \BibitemOpen
  \bibfield  {author} {\bibinfo {author} {\bibfnamefont {E.}~\bibnamefont
  {Torrontegui}}, \bibinfo {author} {\bibfnamefont {S.}~\bibnamefont
  {Ib{\'a}nez}}, \bibinfo {author} {\bibfnamefont {S.}~\bibnamefont
  {Mart{\'\i}nez-Garaot}}, \bibinfo {author} {\bibfnamefont {M.}~\bibnamefont
  {Modugno}}, \bibinfo {author} {\bibfnamefont {A.}~\bibnamefont {del Campo}},
  \bibinfo {author} {\bibfnamefont {D.}~\bibnamefont {Gu{\'e}ry-Odelin}},
  \bibinfo {author} {\bibfnamefont {A.}~\bibnamefont {Ruschhaupt}}, \bibinfo
  {author} {\bibfnamefont {X.}~\bibnamefont {Chen}},\ and\ \bibinfo {author}
  {\bibfnamefont {J.~G.}\ \bibnamefont {Muga}},\ }\bibfield  {title} {\bibinfo
  {title} {Shortcuts to adiabaticity},\ }\href
  {https://doi.org/https://doi.org/10.1016/B978-0-12-408090-4.00002-5}
  {\bibfield  {journal} {\bibinfo  {journal} {Advances in atomic, molecular,
  and optical physics}\ }\textbf {\bibinfo {volume} {62}},\ \bibinfo {pages}
  {117} (\bibinfo {year} {2013})}\BibitemShut {NoStop}%
\bibitem [{\citenamefont {Chen}\ \emph {et~al.}(2010)\citenamefont {Chen},
  \citenamefont {Ruschhaupt}, \citenamefont {Schmidt}, \citenamefont {del
  Campo}, \citenamefont {Gu{\'e}ry-Odelin},\ and\ \citenamefont
  {Muga}}]{chen2010fast}%
  \BibitemOpen
  \bibfield  {author} {\bibinfo {author} {\bibfnamefont {X.}~\bibnamefont
  {Chen}}, \bibinfo {author} {\bibfnamefont {A.}~\bibnamefont {Ruschhaupt}},
  \bibinfo {author} {\bibfnamefont {S.}~\bibnamefont {Schmidt}}, \bibinfo
  {author} {\bibfnamefont {A.}~\bibnamefont {del Campo}}, \bibinfo {author}
  {\bibfnamefont {D.}~\bibnamefont {Gu{\'e}ry-Odelin}},\ and\ \bibinfo {author}
  {\bibfnamefont {J.~G.}\ \bibnamefont {Muga}},\ }\bibfield  {title} {\bibinfo
  {title} {Fast optimal frictionless atom cooling in harmonic traps: Shortcut
  to adiabaticity},\ }\href {https://doi.org/10.1103/PhysRevLett.104.063002}
  {\bibfield  {journal} {\bibinfo  {journal} {Physical Review Letters}\
  }\textbf {\bibinfo {volume} {104}},\ \bibinfo {pages} {063002} (\bibinfo
  {year} {2010})}\BibitemShut {NoStop}%
\bibitem [{\citenamefont {Chen}\ \emph {et~al.}(2011)\citenamefont {Chen},
  \citenamefont {Torrontegui},\ and\ \citenamefont {Muga}}]{chen2011lewis}%
  \BibitemOpen
  \bibfield  {author} {\bibinfo {author} {\bibfnamefont {X.}~\bibnamefont
  {Chen}}, \bibinfo {author} {\bibfnamefont {E.}~\bibnamefont {Torrontegui}},\
  and\ \bibinfo {author} {\bibfnamefont {J.~G.}\ \bibnamefont {Muga}},\
  }\bibfield  {title} {\bibinfo {title} {Lewis-riesenfeld invariants and
  transitionless quantum driving},\ }\href
  {https://doi.org/10.1103/PhysRevA.83.062116} {\bibfield  {journal} {\bibinfo
  {journal} {Physical Review A}\ }\textbf {\bibinfo {volume} {83}},\ \bibinfo
  {pages} {062116} (\bibinfo {year} {2011})}\BibitemShut {NoStop}%
\bibitem [{\citenamefont {Masuda}\ and\ \citenamefont
  {Nakamura}(2008)}]{masuda2008fast}%
  \BibitemOpen
  \bibfield  {author} {\bibinfo {author} {\bibfnamefont {S.}~\bibnamefont
  {Masuda}}\ and\ \bibinfo {author} {\bibfnamefont {K.}~\bibnamefont
  {Nakamura}},\ }\bibfield  {title} {\bibinfo {title} {Fast-forward problem in
  quantum mechanics},\ }\href {https://doi.org/10.1103/PhysRevA.78.062108}
  {\bibfield  {journal} {\bibinfo  {journal} {Physical Review A}\ }\textbf
  {\bibinfo {volume} {78}},\ \bibinfo {pages} {062108} (\bibinfo {year}
  {2008})}\BibitemShut {NoStop}%
\bibitem [{\citenamefont {Masuda}\ and\ \citenamefont
  {Nakamura}(2010)}]{masuda2010fast}%
  \BibitemOpen
  \bibfield  {author} {\bibinfo {author} {\bibfnamefont {S.}~\bibnamefont
  {Masuda}}\ and\ \bibinfo {author} {\bibfnamefont {K.}~\bibnamefont
  {Nakamura}},\ }\bibfield  {title} {\bibinfo {title} {Fast-forward of
  adiabatic dynamics in quantum mechanics},\ }\href
  {https://doi.org/10.1098/rspa.2009.0446} {\bibfield  {journal} {\bibinfo
  {journal} {Proceedings of the Royal Society A: Mathematical, Physical and
  Engineering Sciences}\ }\textbf {\bibinfo {volume} {466}},\ \bibinfo {pages}
  {1135} (\bibinfo {year} {2010})}\BibitemShut {NoStop}%
\bibitem [{\citenamefont {Demirplak}\ and\ \citenamefont
  {Rice}(2003{\natexlab{a}})}]{demirplak2003adiabatic}%
  \BibitemOpen
  \bibfield  {author} {\bibinfo {author} {\bibfnamefont {M.}~\bibnamefont
  {Demirplak}}\ and\ \bibinfo {author} {\bibfnamefont {S.~A.}\ \bibnamefont
  {Rice}},\ }\bibfield  {title} {\bibinfo {title} {Adiabatic population
  transfer with control fields},\ }\href {https://doi.org/10.1021/jp030708a}
  {\bibfield  {journal} {\bibinfo  {journal} {The Journal of Physical Chemistry
  A}\ }\textbf {\bibinfo {volume} {107}},\ \bibinfo {pages} {9937} (\bibinfo
  {year} {2003}{\natexlab{a}})}\BibitemShut {NoStop}%
\bibitem [{\citenamefont {Demirplak}\ and\ \citenamefont
  {Rice}(2005)}]{demirplak2005assisted}%
  \BibitemOpen
  \bibfield  {author} {\bibinfo {author} {\bibfnamefont {M.}~\bibnamefont
  {Demirplak}}\ and\ \bibinfo {author} {\bibfnamefont {S.~A.}\ \bibnamefont
  {Rice}},\ }\bibfield  {title} {\bibinfo {title} {Assisted adiabatic passage
  revisited},\ }\href {https://doi.org/10.1021/jp040647w} {\bibfield  {journal}
  {\bibinfo  {journal} {The Journal of Physical Chemistry B}\ }\textbf
  {\bibinfo {volume} {109}},\ \bibinfo {pages} {6838} (\bibinfo {year}
  {2005})}\BibitemShut {NoStop}%
\bibitem [{\citenamefont {Berry}(2009)}]{berry2009transitionless}%
  \BibitemOpen
  \bibfield  {author} {\bibinfo {author} {\bibfnamefont {M.~V.}\ \bibnamefont
  {Berry}},\ }\bibfield  {title} {\bibinfo {title} {Transitionless quantum
  driving},\ }\href {https://doi.org/10.1088/1751-8113/42/36/365303} {\bibfield
   {journal} {\bibinfo  {journal} {Journal of Physics A: Mathematical and
  Theoretical}\ }\textbf {\bibinfo {volume} {42}},\ \bibinfo {pages} {365303}
  (\bibinfo {year} {2009})}\BibitemShut {NoStop}%
\bibitem [{\citenamefont {del Campo}(2013)}]{PhysRevLett.111.100502}%
  \BibitemOpen
  \bibfield  {author} {\bibinfo {author} {\bibfnamefont {A.}~\bibnamefont {del
  Campo}},\ }\bibfield  {title} {\bibinfo {title} {Shortcuts to adiabaticity by
  counterdiabatic driving},\ }\href
  {https://doi.org/10.1103/PhysRevLett.111.100502} {\bibfield  {journal}
  {\bibinfo  {journal} {Physical Review Letters}\ }\textbf {\bibinfo {volume}
  {111}},\ \bibinfo {pages} {100502} (\bibinfo {year} {2013})}\BibitemShut
  {NoStop}%
\bibitem [{\citenamefont {Takahashi}(2019)}]{takahashi2019hamiltonian}%
  \BibitemOpen
  \bibfield  {author} {\bibinfo {author} {\bibfnamefont {K.}~\bibnamefont
  {Takahashi}},\ }\bibfield  {title} {\bibinfo {title} {Hamiltonian engineering
  for adiabatic quantum computation: Lessons from shortcuts to adiabaticity},\
  }\href {https://doi.org/10.7566/JPSJ.88.061002} {\bibfield  {journal}
  {\bibinfo  {journal} {Journal of the Physical Society of Japan}\ }\textbf
  {\bibinfo {volume} {88}},\ \bibinfo {pages} {061002} (\bibinfo {year}
  {2019})}\BibitemShut {NoStop}%
\bibitem [{\citenamefont {Sels}\ and\ \citenamefont
  {Polkovnikov}(2017)}]{sels2017minimizing}%
  \BibitemOpen
  \bibfield  {author} {\bibinfo {author} {\bibfnamefont {D.}~\bibnamefont
  {Sels}}\ and\ \bibinfo {author} {\bibfnamefont {A.}~\bibnamefont
  {Polkovnikov}},\ }\bibfield  {title} {\bibinfo {title} {Minimizing
  irreversible losses in quantum systems by local counterdiabatic driving},\
  }\href {https://doi.org/10.1073/pnas.1619826114} {\bibfield  {journal}
  {\bibinfo  {journal} {Proceedings of the National Academy of Sciences}\
  }\textbf {\bibinfo {volume} {114}},\ \bibinfo {pages} {E3909} (\bibinfo
  {year} {2017})}\BibitemShut {NoStop}%
\bibitem [{\citenamefont {Claeys}\ \emph {et~al.}(2019)\citenamefont {Claeys},
  \citenamefont {Pandey}, \citenamefont {Sels},\ and\ \citenamefont
  {Polkovnikov}}]{ClaeysPRL}%
  \BibitemOpen
  \bibfield  {author} {\bibinfo {author} {\bibfnamefont {P.~W.}\ \bibnamefont
  {Claeys}}, \bibinfo {author} {\bibfnamefont {M.}~\bibnamefont {Pandey}},
  \bibinfo {author} {\bibfnamefont {D.}~\bibnamefont {Sels}},\ and\ \bibinfo
  {author} {\bibfnamefont {A.}~\bibnamefont {Polkovnikov}},\ }\bibfield
  {title} {\bibinfo {title} {Floquet-engineering counterdiabatic protocols in
  quantum many-body systems},\ }\href
  {https://doi.org/10.1103/PhysRevLett.123.090602} {\bibfield  {journal}
  {\bibinfo  {journal} {Physical Review Letters}\ }\textbf {\bibinfo {volume}
  {123}},\ \bibinfo {pages} {090602} (\bibinfo {year} {2019})}\BibitemShut
  {NoStop}%
\bibitem [{\citenamefont {Hegade}\ \emph
  {et~al.}(2021{\natexlab{a}})\citenamefont {Hegade}, \citenamefont {Paul},
  \citenamefont {Ding}, \citenamefont {Sanz}, \citenamefont
  {Albarr{\'a}n-Arriagada}, \citenamefont {Solano},\ and\ \citenamefont
  {Chen}}]{hegade2021shortcuts}%
  \BibitemOpen
  \bibfield  {author} {\bibinfo {author} {\bibfnamefont {N.~N.}\ \bibnamefont
  {Hegade}}, \bibinfo {author} {\bibfnamefont {K.}~\bibnamefont {Paul}},
  \bibinfo {author} {\bibfnamefont {Y.}~\bibnamefont {Ding}}, \bibinfo {author}
  {\bibfnamefont {M.}~\bibnamefont {Sanz}}, \bibinfo {author} {\bibfnamefont
  {F.}~\bibnamefont {Albarr{\'a}n-Arriagada}}, \bibinfo {author} {\bibfnamefont
  {E.}~\bibnamefont {Solano}},\ and\ \bibinfo {author} {\bibfnamefont
  {X.}~\bibnamefont {Chen}},\ }\bibfield  {title} {\bibinfo {title} {Shortcuts
  to adiabaticity in digitized adiabatic quantum computing},\ }\href
  {https://doi.org/10.1103/PhysRevApplied.15.024038} {\bibfield  {journal}
  {\bibinfo  {journal} {Physical Review Applied}\ }\textbf {\bibinfo {volume}
  {15}},\ \bibinfo {pages} {024038} (\bibinfo {year}
  {2021}{\natexlab{a}})}\BibitemShut {NoStop}%
\bibitem [{\citenamefont {Hegade}\ \emph
  {et~al.}(2021{\natexlab{b}})\citenamefont {Hegade}, \citenamefont {Paul},
  \citenamefont {Albarr{\'a}n-Arriagada}, \citenamefont {Chen},\ and\
  \citenamefont {Solano}}]{hegadePRAletter}%
  \BibitemOpen
  \bibfield  {author} {\bibinfo {author} {\bibfnamefont {N.~N.}\ \bibnamefont
  {Hegade}}, \bibinfo {author} {\bibfnamefont {K.}~\bibnamefont {Paul}},
  \bibinfo {author} {\bibfnamefont {F.}~\bibnamefont {Albarr{\'a}n-Arriagada}},
  \bibinfo {author} {\bibfnamefont {X.}~\bibnamefont {Chen}},\ and\ \bibinfo
  {author} {\bibfnamefont {E.}~\bibnamefont {Solano}},\ }\bibfield  {title}
  {\bibinfo {title} {Digitized adiabatic quantum factorization},\ }\href
  {https://doi.org/10.1103/PhysRevA.104.L050403} {\bibfield  {journal}
  {\bibinfo  {journal} {Physical Review A}\ }\textbf {\bibinfo {volume}
  {104}},\ \bibinfo {pages} {L050403} (\bibinfo {year}
  {2021}{\natexlab{b}})}\BibitemShut {NoStop}%
\bibitem [{\citenamefont {Hegade}\ \emph
  {et~al.}(2021{\natexlab{c}})\citenamefont {Hegade}, \citenamefont
  {Chandarana}, \citenamefont {Paul}, \citenamefont {Chen}, \citenamefont
  {Albarr{\'a}n-Arriagada},\ and\ \citenamefont
  {Solano}}]{hegade2021portfolio}%
  \BibitemOpen
  \bibfield  {author} {\bibinfo {author} {\bibfnamefont {N.}~\bibnamefont
  {Hegade}}, \bibinfo {author} {\bibfnamefont {P.}~\bibnamefont {Chandarana}},
  \bibinfo {author} {\bibfnamefont {K.}~\bibnamefont {Paul}}, \bibinfo {author}
  {\bibfnamefont {X.}~\bibnamefont {Chen}}, \bibinfo {author} {\bibfnamefont
  {F.}~\bibnamefont {Albarr{\'a}n-Arriagada}},\ and\ \bibinfo {author}
  {\bibfnamefont {E.}~\bibnamefont {Solano}},\ }\bibfield  {title} {\bibinfo
  {title} {Portfolio optimization with digitized-counterdiabatic quantum
  algorithms},\ }\href {https://arxiv.org/abs/2112.08347} {\bibfield  {journal}
  {\bibinfo  {journal} {arXiv preprint arXiv:2112.08347}\ } (\bibinfo {year}
  {2021}{\natexlab{c}})}\BibitemShut {NoStop}%
\bibitem [{\citenamefont {Opatrn{\`y}}\ and\ \citenamefont
  {M{\o}lmer}(2014)}]{opatrny2014partial}%
  \BibitemOpen
  \bibfield  {author} {\bibinfo {author} {\bibfnamefont {T.}~\bibnamefont
  {Opatrn{\`y}}}\ and\ \bibinfo {author} {\bibfnamefont {K.}~\bibnamefont
  {M{\o}lmer}},\ }\bibfield  {title} {\bibinfo {title} {Partial suppression of
  nonadiabatic transitions},\ }\href
  {https://doi.org/10.1088/1367-2630/16/1/015025} {\bibfield  {journal}
  {\bibinfo  {journal} {New Journal of Physics}\ }\textbf {\bibinfo {volume}
  {16}},\ \bibinfo {pages} {015025} (\bibinfo {year} {2014})}\BibitemShut
  {NoStop}%
\bibitem [{\citenamefont {Petiziol}\ \emph {et~al.}(2019)\citenamefont
  {Petiziol}, \citenamefont {Dive}, \citenamefont {Carretta}, \citenamefont
  {Mannella}, \citenamefont {Mintert},\ and\ \citenamefont
  {Wimberger}}]{petiziol2019accelerating}%
  \BibitemOpen
  \bibfield  {author} {\bibinfo {author} {\bibfnamefont {F.}~\bibnamefont
  {Petiziol}}, \bibinfo {author} {\bibfnamefont {B.}~\bibnamefont {Dive}},
  \bibinfo {author} {\bibfnamefont {S.}~\bibnamefont {Carretta}}, \bibinfo
  {author} {\bibfnamefont {R.}~\bibnamefont {Mannella}}, \bibinfo {author}
  {\bibfnamefont {F.}~\bibnamefont {Mintert}},\ and\ \bibinfo {author}
  {\bibfnamefont {S.}~\bibnamefont {Wimberger}},\ }\bibfield  {title} {\bibinfo
  {title} {Accelerating adiabatic protocols for entangling two qubits in
  circuit qed},\ }\href {https://doi.org/10.1103/PhysRevA.99.042315} {\bibfield
   {journal} {\bibinfo  {journal} {Physical Review A}\ }\textbf {\bibinfo
  {volume} {99}},\ \bibinfo {pages} {042315} (\bibinfo {year}
  {2019})}\BibitemShut {NoStop}%
\bibitem [{\citenamefont {Zhou}\ \emph
  {et~al.}(2020{\natexlab{b}})\citenamefont {Zhou}, \citenamefont {Ji},
  \citenamefont {Nie}, \citenamefont {Yang}, \citenamefont {Chen},
  \citenamefont {Bian},\ and\ \citenamefont {Peng}}]{zhou2020experimental}%
  \BibitemOpen
  \bibfield  {author} {\bibinfo {author} {\bibfnamefont {H.}~\bibnamefont
  {Zhou}}, \bibinfo {author} {\bibfnamefont {Y.}~\bibnamefont {Ji}}, \bibinfo
  {author} {\bibfnamefont {X.}~\bibnamefont {Nie}}, \bibinfo {author}
  {\bibfnamefont {X.}~\bibnamefont {Yang}}, \bibinfo {author} {\bibfnamefont
  {X.}~\bibnamefont {Chen}}, \bibinfo {author} {\bibfnamefont {J.}~\bibnamefont
  {Bian}},\ and\ \bibinfo {author} {\bibfnamefont {X.}~\bibnamefont {Peng}},\
  }\bibfield  {title} {\bibinfo {title} {Experimental realization of shortcuts
  to adiabaticity in a nonintegrable spin chain by local counterdiabatic
  driving},\ }\href {https://doi.org/10.1103/PhysRevApplied.13.044059}
  {\bibfield  {journal} {\bibinfo  {journal} {Physical Review Applied}\
  }\textbf {\bibinfo {volume} {13}},\ \bibinfo {pages} {044059} (\bibinfo
  {year} {2020}{\natexlab{b}})}\BibitemShut {NoStop}%
\bibitem [{\citenamefont {Ji}\ \emph {et~al.}(2019)\citenamefont {Ji},
  \citenamefont {Bian}, \citenamefont {Chen}, \citenamefont {Li}, \citenamefont
  {Nie}, \citenamefont {Zhou},\ and\ \citenamefont
  {Peng}}]{ji2019experimental}%
  \BibitemOpen
  \bibfield  {author} {\bibinfo {author} {\bibfnamefont {Y.}~\bibnamefont
  {Ji}}, \bibinfo {author} {\bibfnamefont {J.}~\bibnamefont {Bian}}, \bibinfo
  {author} {\bibfnamefont {X.}~\bibnamefont {Chen}}, \bibinfo {author}
  {\bibfnamefont {J.}~\bibnamefont {Li}}, \bibinfo {author} {\bibfnamefont
  {X.}~\bibnamefont {Nie}}, \bibinfo {author} {\bibfnamefont {H.}~\bibnamefont
  {Zhou}},\ and\ \bibinfo {author} {\bibfnamefont {X.}~\bibnamefont {Peng}},\
  }\bibfield  {title} {\bibinfo {title} {Experimental preparation of
  greenberger-horne-zeilinger states in an ising spin model by partially
  suppressing the nonadiabatic transitions},\ }\href
  {https://doi.org/10.1103/PhysRevA.99.032323} {\bibfield  {journal} {\bibinfo
  {journal} {Physical Review A}\ }\textbf {\bibinfo {volume} {99}},\ \bibinfo
  {pages} {032323} (\bibinfo {year} {2019})}\BibitemShut {NoStop}%
\bibitem [{\citenamefont {Passarelli}\ \emph {et~al.}(2020)\citenamefont
  {Passarelli}, \citenamefont {Cataudella}, \citenamefont {Fazio},\ and\
  \citenamefont {Lucignano}}]{passarelli2020counterdiabatic}%
  \BibitemOpen
  \bibfield  {author} {\bibinfo {author} {\bibfnamefont {G.}~\bibnamefont
  {Passarelli}}, \bibinfo {author} {\bibfnamefont {V.}~\bibnamefont
  {Cataudella}}, \bibinfo {author} {\bibfnamefont {R.}~\bibnamefont {Fazio}},\
  and\ \bibinfo {author} {\bibfnamefont {P.}~\bibnamefont {Lucignano}},\
  }\bibfield  {title} {\bibinfo {title} {Counterdiabatic driving in the quantum
  annealing of the p-spin model: A variational approach},\ }\href
  {https://doi.org/10.1103/PhysRevResearch.2.013283} {\bibfield  {journal}
  {\bibinfo  {journal} {Physical Review Research}\ }\textbf {\bibinfo {volume}
  {2}},\ \bibinfo {pages} {013283} (\bibinfo {year} {2020})}\BibitemShut
  {NoStop}%
\bibitem [{\citenamefont {Takahashi}(2017)}]{takahashi2017shortcuts}%
  \BibitemOpen
  \bibfield  {author} {\bibinfo {author} {\bibfnamefont {K.}~\bibnamefont
  {Takahashi}},\ }\bibfield  {title} {\bibinfo {title} {Shortcuts to
  adiabaticity for quantum annealing},\ }\href
  {https://doi.org/10.1103/PhysRevA.95.012309} {\bibfield  {journal} {\bibinfo
  {journal} {Physical Review A}\ }\textbf {\bibinfo {volume} {95}},\ \bibinfo
  {pages} {012309} (\bibinfo {year} {2017})}\BibitemShut {NoStop}%
\bibitem [{\citenamefont {Vinci}\ and\ \citenamefont
  {Lidar}(2017)}]{vinci2017non}%
  \BibitemOpen
  \bibfield  {author} {\bibinfo {author} {\bibfnamefont {W.}~\bibnamefont
  {Vinci}}\ and\ \bibinfo {author} {\bibfnamefont {D.~A.}\ \bibnamefont
  {Lidar}},\ }\bibfield  {title} {\bibinfo {title} {Non-stoquastic hamiltonians
  in quantum annealing via geometric phases},\ }\href
  {https://doi.org/10.1038/s41534-017-0037-z} {\bibfield  {journal} {\bibinfo
  {journal} {npj Quantum Information}\ }\textbf {\bibinfo {volume} {3}},\
  \bibinfo {pages} {1} (\bibinfo {year} {2017})}\BibitemShut {NoStop}%
\bibitem [{\citenamefont {Hegade}\ \emph {et~al.}(2022)\citenamefont {Hegade},
  \citenamefont {Chen},\ and\ \citenamefont {Solano}}]{hegade2022digitized}%
  \BibitemOpen
  \bibfield  {author} {\bibinfo {author} {\bibfnamefont {N.~N.}\ \bibnamefont
  {Hegade}}, \bibinfo {author} {\bibfnamefont {X.}~\bibnamefont {Chen}},\ and\
  \bibinfo {author} {\bibfnamefont {E.}~\bibnamefont {Solano}},\ }\bibfield
  {title} {\bibinfo {title} {Digitized-counterdiabatic quantum optimization},\
  }\href {https://arxiv.org/abs/2201.00790} {\bibfield  {journal} {\bibinfo
  {journal} {arXiv preprint arXiv:2201.00790}\ } (\bibinfo {year}
  {2022})}\BibitemShut {NoStop}%
\bibitem [{\citenamefont {Yao}\ \emph {et~al.}(2021)\citenamefont {Yao},
  \citenamefont {Lin},\ and\ \citenamefont {Bukov}}]{yao2021reinforcement}%
  \BibitemOpen
  \bibfield  {author} {\bibinfo {author} {\bibfnamefont {J.}~\bibnamefont
  {Yao}}, \bibinfo {author} {\bibfnamefont {L.}~\bibnamefont {Lin}},\ and\
  \bibinfo {author} {\bibfnamefont {M.}~\bibnamefont {Bukov}},\ }\bibfield
  {title} {\bibinfo {title} {Reinforcement learning for many-body ground-state
  preparation inspired by counterdiabatic driving},\ }\href
  {https://doi.org/10.1103/PhysRevX.11.031070} {\bibfield  {journal} {\bibinfo
  {journal} {Physical Review X}\ }\textbf {\bibinfo {volume} {11}},\ \bibinfo
  {pages} {031070} (\bibinfo {year} {2021})}\BibitemShut {NoStop}%
\bibitem [{\citenamefont {Chandarana}\ \emph {et~al.}(2022)\citenamefont
  {Chandarana}, \citenamefont {Hegade}, \citenamefont {Paul}, \citenamefont
  {Albarr{\'a}n-Arriagada}, \citenamefont {Solano}, \citenamefont {del Campo},\
  and\ \citenamefont {Chen}}]{chandarana2022digitized}%
  \BibitemOpen
  \bibfield  {author} {\bibinfo {author} {\bibfnamefont {P.}~\bibnamefont
  {Chandarana}}, \bibinfo {author} {\bibfnamefont {N.~N.}\ \bibnamefont
  {Hegade}}, \bibinfo {author} {\bibfnamefont {K.}~\bibnamefont {Paul}},
  \bibinfo {author} {\bibfnamefont {F.}~\bibnamefont {Albarr{\'a}n-Arriagada}},
  \bibinfo {author} {\bibfnamefont {E.}~\bibnamefont {Solano}}, \bibinfo
  {author} {\bibfnamefont {A.}~\bibnamefont {del Campo}},\ and\ \bibinfo
  {author} {\bibfnamefont {X.}~\bibnamefont {Chen}},\ }\bibfield  {title}
  {\bibinfo {title} {Digitized-counterdiabatic quantum approximate optimization
  algorithm},\ }\href {https://doi.org/10.1103/PhysRevResearch.4.013141}
  {\bibfield  {journal} {\bibinfo  {journal} {Physical Review Research}\
  }\textbf {\bibinfo {volume} {4}},\ \bibinfo {pages} {013141} (\bibinfo {year}
  {2022})}\BibitemShut {NoStop}%
\bibitem [{\citenamefont {Wurtz}\ and\ \citenamefont
  {Love}(2022)}]{wurtz2022counterdiabaticity}%
  \BibitemOpen
  \bibfield  {author} {\bibinfo {author} {\bibfnamefont {J.}~\bibnamefont
  {Wurtz}}\ and\ \bibinfo {author} {\bibfnamefont {P.~J.}\ \bibnamefont
  {Love}},\ }\bibfield  {title} {\bibinfo {title} {Counterdiabaticity and the
  quantum approximate optimization algorithm},\ }\href
  {https://doi.org/10.22331/q-2022-01-27-635} {\bibfield  {journal} {\bibinfo
  {journal} {Quantum}\ }\textbf {\bibinfo {volume} {6}},\ \bibinfo {pages}
  {635} (\bibinfo {year} {2022})}\BibitemShut {NoStop}%
\bibitem [{\citenamefont {Chai}\ \emph {et~al.}(2022)\citenamefont {Chai},
  \citenamefont {Han}, \citenamefont {Wu}, \citenamefont {Li}, \citenamefont
  {Dou},\ and\ \citenamefont {Guo}}]{PhysRevA.105.042415}%
  \BibitemOpen
  \bibfield  {author} {\bibinfo {author} {\bibfnamefont {Y.}~\bibnamefont
  {Chai}}, \bibinfo {author} {\bibfnamefont {Y.-J.}\ \bibnamefont {Han}},
  \bibinfo {author} {\bibfnamefont {Y.-C.}\ \bibnamefont {Wu}}, \bibinfo
  {author} {\bibfnamefont {Y.}~\bibnamefont {Li}}, \bibinfo {author}
  {\bibfnamefont {M.}~\bibnamefont {Dou}},\ and\ \bibinfo {author}
  {\bibfnamefont {G.-P.}\ \bibnamefont {Guo}},\ }\bibfield  {title} {\bibinfo
  {title} {Shortcuts to the quantum approximate optimization algorithm},\
  }\href {https://doi.org/10.1103/PhysRevA.105.042415} {\bibfield  {journal}
  {\bibinfo  {journal} {Physical Review A}\ }\textbf {\bibinfo {volume}
  {105}},\ \bibinfo {pages} {042415} (\bibinfo {year} {2022})}\BibitemShut
  {NoStop}%
\bibitem [{\citenamefont {Schuld}\ \emph {et~al.}(2019)\citenamefont {Schuld},
  \citenamefont {Bergholm}, \citenamefont {Gogolin}, \citenamefont {Izaac},\
  and\ \citenamefont {Killoran}}]{schuld2019evaluating}%
  \BibitemOpen
  \bibfield  {author} {\bibinfo {author} {\bibfnamefont {M.}~\bibnamefont
  {Schuld}}, \bibinfo {author} {\bibfnamefont {V.}~\bibnamefont {Bergholm}},
  \bibinfo {author} {\bibfnamefont {C.}~\bibnamefont {Gogolin}}, \bibinfo
  {author} {\bibfnamefont {J.}~\bibnamefont {Izaac}},\ and\ \bibinfo {author}
  {\bibfnamefont {N.}~\bibnamefont {Killoran}},\ }\bibfield  {title} {\bibinfo
  {title} {Evaluating analytic gradients on quantum hardware},\ }\href
  {https://doi.org/10.1103/PhysRevA.99.032331} {\bibfield  {journal} {\bibinfo
  {journal} {Physical Review A}\ }\textbf {\bibinfo {volume} {99}},\ \bibinfo
  {pages} {032331} (\bibinfo {year} {2019})}\BibitemShut {NoStop}%
\bibitem [{\citenamefont {Verdon}\ \emph {et~al.}(2018)\citenamefont {Verdon},
  \citenamefont {Pye},\ and\ \citenamefont {Broughton}}]{verdon2018universal}%
  \BibitemOpen
  \bibfield  {author} {\bibinfo {author} {\bibfnamefont {G.}~\bibnamefont
  {Verdon}}, \bibinfo {author} {\bibfnamefont {J.}~\bibnamefont {Pye}},\ and\
  \bibinfo {author} {\bibfnamefont {M.}~\bibnamefont {Broughton}},\ }\bibfield
  {title} {\bibinfo {title} {A universal training algorithm for quantum deep
  learning},\ }\href {https://arxiv.org/abs/1806.09729} {\bibfield  {journal}
  {\bibinfo  {journal} {arXiv preprint arXiv:1806.09729}\ } (\bibinfo {year}
  {2018})}\BibitemShut {NoStop}%
\bibitem [{\citenamefont {Farhi}\ and\ \citenamefont
  {Neven}(2018)}]{farhi2018classification}%
  \BibitemOpen
  \bibfield  {author} {\bibinfo {author} {\bibfnamefont {E.}~\bibnamefont
  {Farhi}}\ and\ \bibinfo {author} {\bibfnamefont {H.}~\bibnamefont {Neven}},\
  }\bibfield  {title} {\bibinfo {title} {Classification with quantum neural
  networks on near term processors},\ }\href {https://arxiv.org/abs/1802.06002}
  {\bibfield  {journal} {\bibinfo  {journal} {arXiv preprint arXiv:1802.06002}\
  } (\bibinfo {year} {2018})}\BibitemShut {NoStop}%
\bibitem [{\citenamefont {Chen}\ \emph {et~al.}(2021)\citenamefont {Chen},
  \citenamefont {Wossnig}, \citenamefont {Severini}, \citenamefont {Neven},\
  and\ \citenamefont {Mohseni}}]{chen2021universal}%
  \BibitemOpen
  \bibfield  {author} {\bibinfo {author} {\bibfnamefont {H.}~\bibnamefont
  {Chen}}, \bibinfo {author} {\bibfnamefont {L.}~\bibnamefont {Wossnig}},
  \bibinfo {author} {\bibfnamefont {S.}~\bibnamefont {Severini}}, \bibinfo
  {author} {\bibfnamefont {H.}~\bibnamefont {Neven}},\ and\ \bibinfo {author}
  {\bibfnamefont {M.}~\bibnamefont {Mohseni}},\ }\bibfield  {title} {\bibinfo
  {title} {Universal discriminative quantum neural networks},\ }\href
  {https://doi.org/10.1007/s42484-020-00025-7} {\bibfield  {journal} {\bibinfo
  {journal} {Quantum Machine Intelligence}\ }\textbf {\bibinfo {volume} {3}},\
  \bibinfo {pages} {1} (\bibinfo {year} {2021})}\BibitemShut {NoStop}%
\bibitem [{\citenamefont {Yang}\ \emph {et~al.}(2017)\citenamefont {Yang},
  \citenamefont {Rahmani}, \citenamefont {Shabani}, \citenamefont {Neven},\
  and\ \citenamefont {Chamon}}]{yang2017optimizing}%
  \BibitemOpen
  \bibfield  {author} {\bibinfo {author} {\bibfnamefont {Z.-C.}\ \bibnamefont
  {Yang}}, \bibinfo {author} {\bibfnamefont {A.}~\bibnamefont {Rahmani}},
  \bibinfo {author} {\bibfnamefont {A.}~\bibnamefont {Shabani}}, \bibinfo
  {author} {\bibfnamefont {H.}~\bibnamefont {Neven}},\ and\ \bibinfo {author}
  {\bibfnamefont {C.}~\bibnamefont {Chamon}},\ }\bibfield  {title} {\bibinfo
  {title} {Optimizing variational quantum algorithms using pontryagin’s
  minimum principle},\ }\href {https://doi.org/10.1103/PhysRevX.7.021027}
  {\bibfield  {journal} {\bibinfo  {journal} {Physical Review X}\ }\textbf
  {\bibinfo {volume} {7}},\ \bibinfo {pages} {021027} (\bibinfo {year}
  {2017})}\BibitemShut {NoStop}%
\bibitem [{\citenamefont {Grant}\ \emph {et~al.}(2019)\citenamefont {Grant},
  \citenamefont {Wossnig}, \citenamefont {Ostaszewski},\ and\ \citenamefont
  {Benedetti}}]{grant2019initialization}%
  \BibitemOpen
  \bibfield  {author} {\bibinfo {author} {\bibfnamefont {E.}~\bibnamefont
  {Grant}}, \bibinfo {author} {\bibfnamefont {L.}~\bibnamefont {Wossnig}},
  \bibinfo {author} {\bibfnamefont {M.}~\bibnamefont {Ostaszewski}},\ and\
  \bibinfo {author} {\bibfnamefont {M.}~\bibnamefont {Benedetti}},\ }\bibfield
  {title} {\bibinfo {title} {An initialization strategy for addressing barren
  plateaus in parametrized quantum circuits},\ }\href
  {https://doi.org/10.22331/q-2019-12-09-214} {\bibfield  {journal} {\bibinfo
  {journal} {Quantum}\ }\textbf {\bibinfo {volume} {3}},\ \bibinfo {pages}
  {214} (\bibinfo {year} {2019})}\BibitemShut {NoStop}%
\bibitem [{\citenamefont {Egger}\ \emph {et~al.}(2021)\citenamefont {Egger},
  \citenamefont {Mare{\v{c}}ek},\ and\ \citenamefont
  {Woerner}}]{Egger2021warmstartingquantum}%
  \BibitemOpen
  \bibfield  {author} {\bibinfo {author} {\bibfnamefont {D.~J.}\ \bibnamefont
  {Egger}}, \bibinfo {author} {\bibfnamefont {J.}~\bibnamefont
  {Mare{\v{c}}ek}},\ and\ \bibinfo {author} {\bibfnamefont {S.}~\bibnamefont
  {Woerner}},\ }\bibfield  {title} {\bibinfo {title} {Warm-starting quantum
  optimization},\ }\href {https://doi.org/10.22331/q-2021-06-17-479} {\bibfield
   {journal} {\bibinfo  {journal} {{Quantum}}\ }\textbf {\bibinfo {volume}
  {5}},\ \bibinfo {pages} {479} (\bibinfo {year} {2021})}\BibitemShut {NoStop}%
\bibitem [{\citenamefont {Sack}\ and\ \citenamefont
  {Serbyn}(2021)}]{Sack2021quantumannealing}%
  \BibitemOpen
  \bibfield  {author} {\bibinfo {author} {\bibfnamefont {S.~H.}\ \bibnamefont
  {Sack}}\ and\ \bibinfo {author} {\bibfnamefont {M.}~\bibnamefont {Serbyn}},\
  }\bibfield  {title} {\bibinfo {title} {Quantum annealing initialization of
  the quantum approximate optimization algorithm},\ }\href
  {https://doi.org/10.22331/q-2021-07-01-491} {\bibfield  {journal} {\bibinfo
  {journal} {{Quantum}}\ }\textbf {\bibinfo {volume} {5}},\ \bibinfo {pages}
  {491} (\bibinfo {year} {2021})}\BibitemShut {NoStop}%
\bibitem [{\citenamefont {Sutton}\ and\ \citenamefont
  {Barto}(2018)}]{sutton2018reinforcement}%
  \BibitemOpen
  \bibfield  {author} {\bibinfo {author} {\bibfnamefont {R.}~\bibnamefont
  {Sutton}}\ and\ \bibinfo {author} {\bibfnamefont {A.}~\bibnamefont {Barto}},\
  }\bibfield  {title} {\bibinfo {title} {Reinforcement learning, second
  edition: An introduction},\ }\href
  {https://books.google.co.in/books?id=uWV0DwAAQBAJ} {\bibfield  {journal}
  {\bibinfo  {journal} {Adaptive Computation and Machine Learning series}\ }
  (\bibinfo {year} {2018})}\BibitemShut {NoStop}%
\bibitem [{\citenamefont {Rose}\ \emph {et~al.}(2021)\citenamefont {Rose},
  \citenamefont {Mair},\ and\ \citenamefont
  {Garrahan}}]{rose2021reinforcement}%
  \BibitemOpen
  \bibfield  {author} {\bibinfo {author} {\bibfnamefont {D.~C.}\ \bibnamefont
  {Rose}}, \bibinfo {author} {\bibfnamefont {J.~F.}\ \bibnamefont {Mair}},\
  and\ \bibinfo {author} {\bibfnamefont {J.~P.}\ \bibnamefont {Garrahan}},\
  }\bibfield  {title} {\bibinfo {title} {A reinforcement learning approach to
  rare trajectory sampling},\ }\href {https://doi.org/10.1088/1367-2630/abd7bd}
  {\bibfield  {journal} {\bibinfo  {journal} {New Journal of Physics}\ }\textbf
  {\bibinfo {volume} {23}},\ \bibinfo {pages} {013013} (\bibinfo {year}
  {2021})}\BibitemShut {NoStop}%
\bibitem [{\citenamefont {Bukov}(2018)}]{bukov2018reinforcement1}%
  \BibitemOpen
  \bibfield  {author} {\bibinfo {author} {\bibfnamefont {M.}~\bibnamefont
  {Bukov}},\ }\bibfield  {title} {\bibinfo {title} {Reinforcement learning for
  autonomous preparation of floquet-engineered states: Inverting the quantum
  kapitza oscillator},\ }\href {https://doi.org/10.1103/PhysRevB.98.224305}
  {\bibfield  {journal} {\bibinfo  {journal} {Physical Review B}\ }\textbf
  {\bibinfo {volume} {98}},\ \bibinfo {pages} {224305} (\bibinfo {year}
  {2018})}\BibitemShut {NoStop}%
\bibitem [{\citenamefont {Bukov}\ \emph {et~al.}(2018)\citenamefont {Bukov},
  \citenamefont {Day}, \citenamefont {Sels}, \citenamefont {Weinberg},
  \citenamefont {Polkovnikov},\ and\ \citenamefont
  {Mehta}}]{bukov2018reinforcement2}%
  \BibitemOpen
  \bibfield  {author} {\bibinfo {author} {\bibfnamefont {M.}~\bibnamefont
  {Bukov}}, \bibinfo {author} {\bibfnamefont {A.~G.}\ \bibnamefont {Day}},
  \bibinfo {author} {\bibfnamefont {D.}~\bibnamefont {Sels}}, \bibinfo {author}
  {\bibfnamefont {P.}~\bibnamefont {Weinberg}}, \bibinfo {author}
  {\bibfnamefont {A.}~\bibnamefont {Polkovnikov}},\ and\ \bibinfo {author}
  {\bibfnamefont {P.}~\bibnamefont {Mehta}},\ }\bibfield  {title} {\bibinfo
  {title} {Reinforcement learning in different phases of quantum control},\
  }\href {https://doi.org/10.1103/PhysRevX.8.031086} {\bibfield  {journal}
  {\bibinfo  {journal} {Physical Review X}\ }\textbf {\bibinfo {volume} {8}},\
  \bibinfo {pages} {031086} (\bibinfo {year} {2018})}\BibitemShut {NoStop}%
\bibitem [{\citenamefont {Niu}\ \emph {et~al.}(2019)\citenamefont {Niu},
  \citenamefont {Boixo}, \citenamefont {Smelyanskiy},\ and\ \citenamefont
  {Neven}}]{niu2019universal}%
  \BibitemOpen
  \bibfield  {author} {\bibinfo {author} {\bibfnamefont {M.~Y.}\ \bibnamefont
  {Niu}}, \bibinfo {author} {\bibfnamefont {S.}~\bibnamefont {Boixo}}, \bibinfo
  {author} {\bibfnamefont {V.~N.}\ \bibnamefont {Smelyanskiy}},\ and\ \bibinfo
  {author} {\bibfnamefont {H.}~\bibnamefont {Neven}},\ }\bibfield  {title}
  {\bibinfo {title} {Universal quantum control through deep reinforcement
  learning},\ }\href {https://doi.org/10.1038/s41534-019-0141-3} {\bibfield
  {journal} {\bibinfo  {journal} {npj Quantum Information}\ }\textbf {\bibinfo
  {volume} {5}},\ \bibinfo {pages} {1} (\bibinfo {year} {2019})}\BibitemShut
  {NoStop}%
\bibitem [{\citenamefont {August}\ and\ \citenamefont
  {Hern{\'a}ndez-Lobato}(2018)}]{august2018taking}%
  \BibitemOpen
  \bibfield  {author} {\bibinfo {author} {\bibfnamefont {M.}~\bibnamefont
  {August}}\ and\ \bibinfo {author} {\bibfnamefont {J.~M.}\ \bibnamefont
  {Hern{\'a}ndez-Lobato}},\ }\bibfield  {title} {\bibinfo {title} {Taking
  gradients through experiments: Lstms and memory proximal policy optimization
  for black-box quantum control},\ }\href
  {https://link.springer.com/chapter/10.1007/978-3-030-02465-9_43} {\bibfield
  {journal} {\bibinfo  {journal} {International Conference on High Performance
  Computing}\ ,\ \bibinfo {pages} {591}} (\bibinfo {year} {2018})}\BibitemShut
  {NoStop}%
\bibitem [{\citenamefont {Porotti}\ \emph {et~al.}(2019)\citenamefont
  {Porotti}, \citenamefont {Tamascelli}, \citenamefont {Restelli},\ and\
  \citenamefont {Prati}}]{porotti2019coherent}%
  \BibitemOpen
  \bibfield  {author} {\bibinfo {author} {\bibfnamefont {R.}~\bibnamefont
  {Porotti}}, \bibinfo {author} {\bibfnamefont {D.}~\bibnamefont {Tamascelli}},
  \bibinfo {author} {\bibfnamefont {M.}~\bibnamefont {Restelli}},\ and\
  \bibinfo {author} {\bibfnamefont {E.}~\bibnamefont {Prati}},\ }\bibfield
  {title} {\bibinfo {title} {Coherent transport of quantum states by deep
  reinforcement learning},\ }\href {https://doi.org/10.1038/s42005-019-0169-x}
  {\bibfield  {journal} {\bibinfo  {journal} {Communications Physics}\ }\textbf
  {\bibinfo {volume} {2}},\ \bibinfo {pages} {1} (\bibinfo {year}
  {2019})}\BibitemShut {NoStop}%
\bibitem [{\citenamefont {Dalgaard}\ \emph {et~al.}(2020)\citenamefont
  {Dalgaard}, \citenamefont {Motzoi}, \citenamefont {S{\o}rensen},\ and\
  \citenamefont {Sherson}}]{dalgaard2020global}%
  \BibitemOpen
  \bibfield  {author} {\bibinfo {author} {\bibfnamefont {M.}~\bibnamefont
  {Dalgaard}}, \bibinfo {author} {\bibfnamefont {F.}~\bibnamefont {Motzoi}},
  \bibinfo {author} {\bibfnamefont {J.~J.}\ \bibnamefont {S{\o}rensen}},\ and\
  \bibinfo {author} {\bibfnamefont {J.}~\bibnamefont {Sherson}},\ }\bibfield
  {title} {\bibinfo {title} {Global optimization of quantum dynamics with
  alphazero deep exploration},\ }\href
  {https://doi.org/10.1038/s41534-019-0241-0} {\bibfield  {journal} {\bibinfo
  {journal} {npj Quantum Information}\ }\textbf {\bibinfo {volume} {6}},\
  \bibinfo {pages} {1} (\bibinfo {year} {2020})}\BibitemShut {NoStop}%
\bibitem [{\citenamefont {Yao}\ \emph {et~al.}(2020)\citenamefont {Yao},
  \citenamefont {Bukov},\ and\ \citenamefont {Lin}}]{yao2020policy}%
  \BibitemOpen
  \bibfield  {author} {\bibinfo {author} {\bibfnamefont {J.}~\bibnamefont
  {Yao}}, \bibinfo {author} {\bibfnamefont {M.}~\bibnamefont {Bukov}},\ and\
  \bibinfo {author} {\bibfnamefont {L.}~\bibnamefont {Lin}},\ }\bibfield
  {title} {\bibinfo {title} {Policy gradient based quantum approximate
  optimization algorithm},\ }\href
  {https://proceedings.mlr.press/v107/yao20a.html} {\bibfield  {journal}
  {\bibinfo  {journal} {Mathematical and Scientific Machine Learning}\ ,\
  \bibinfo {pages} {605}} (\bibinfo {year} {2020})}\BibitemShut {NoStop}%
\bibitem [{\citenamefont {Garcia-Saez}\ and\ \citenamefont
  {Riu}(2019)}]{garcia2019quantum}%
  \BibitemOpen
  \bibfield  {author} {\bibinfo {author} {\bibfnamefont {A.}~\bibnamefont
  {Garcia-Saez}}\ and\ \bibinfo {author} {\bibfnamefont {J.}~\bibnamefont
  {Riu}},\ }\bibfield  {title} {\bibinfo {title} {Quantum observables for
  continuous control of the quantum approximate optimization algorithm via
  reinforcement learning},\ }\href {https://arxiv.org/abs/1911.09682}
  {\bibfield  {journal} {\bibinfo  {journal} {arXiv preprint arXiv:1911.09682}\
  } (\bibinfo {year} {2019})}\BibitemShut {NoStop}%
\bibitem [{\citenamefont {Wauters}\ \emph {et~al.}(2020)\citenamefont
  {Wauters}, \citenamefont {Panizon}, \citenamefont {Mbeng},\ and\
  \citenamefont {Santoro}}]{wauters2020reinforcement}%
  \BibitemOpen
  \bibfield  {author} {\bibinfo {author} {\bibfnamefont {M.~M.}\ \bibnamefont
  {Wauters}}, \bibinfo {author} {\bibfnamefont {E.}~\bibnamefont {Panizon}},
  \bibinfo {author} {\bibfnamefont {G.~B.}\ \bibnamefont {Mbeng}},\ and\
  \bibinfo {author} {\bibfnamefont {G.~E.}\ \bibnamefont {Santoro}},\
  }\bibfield  {title} {\bibinfo {title} {Reinforcement-learning-assisted
  quantum optimization},\ }\href
  {https://doi.org/10.1103/PhysRevResearch.2.033446} {\bibfield  {journal}
  {\bibinfo  {journal} {Physical Review Research}\ }\textbf {\bibinfo {volume}
  {2}},\ \bibinfo {pages} {033446} (\bibinfo {year} {2020})}\BibitemShut
  {NoStop}%
\bibitem [{\citenamefont {Yao}\ \emph {et~al.}(2022)\citenamefont {Yao},
  \citenamefont {Li}, \citenamefont {Bukov}, \citenamefont {Lin},\ and\
  \citenamefont {Ying}}]{yao2022monte}%
  \BibitemOpen
  \bibfield  {author} {\bibinfo {author} {\bibfnamefont {J.}~\bibnamefont
  {Yao}}, \bibinfo {author} {\bibfnamefont {H.}~\bibnamefont {Li}}, \bibinfo
  {author} {\bibfnamefont {M.}~\bibnamefont {Bukov}}, \bibinfo {author}
  {\bibfnamefont {L.}~\bibnamefont {Lin}},\ and\ \bibinfo {author}
  {\bibfnamefont {L.}~\bibnamefont {Ying}},\ }\bibfield  {title} {\bibinfo
  {title} {Monte carlo tree search based hybrid optimization of variational
  quantum circuits},\ }\href {https://arxiv.org/abs/2203.16707} {\bibfield
  {journal} {\bibinfo  {journal} {arXiv preprint arXiv:2203.16707}\ } (\bibinfo
  {year} {2022})}\BibitemShut {NoStop}%
\bibitem [{\citenamefont {Chen}\ \emph {et~al.}(2017)\citenamefont {Chen},
  \citenamefont {Hoffman}, \citenamefont {Colmenarejo}, \citenamefont {Denil},
  \citenamefont {Lillicrap}, \citenamefont {Botvinick},\ and\ \citenamefont
  {Freitas}}]{chen2017learning}%
  \BibitemOpen
  \bibfield  {author} {\bibinfo {author} {\bibfnamefont {Y.}~\bibnamefont
  {Chen}}, \bibinfo {author} {\bibfnamefont {M.~W.}\ \bibnamefont {Hoffman}},
  \bibinfo {author} {\bibfnamefont {S.~G.}\ \bibnamefont {Colmenarejo}},
  \bibinfo {author} {\bibfnamefont {M.}~\bibnamefont {Denil}}, \bibinfo
  {author} {\bibfnamefont {T.~P.}\ \bibnamefont {Lillicrap}}, \bibinfo {author}
  {\bibfnamefont {M.}~\bibnamefont {Botvinick}},\ and\ \bibinfo {author}
  {\bibfnamefont {N.}~\bibnamefont {Freitas}},\ }\bibfield  {title} {\bibinfo
  {title} {Learning to learn without gradient descent by gradient descent},\
  }\href {https://arxiv.org/abs/1611.03824} {\bibfield  {journal} {\bibinfo
  {journal} {International Conference on Machine Learning}\ ,\ \bibinfo {pages}
  {748}} (\bibinfo {year} {2017})}\BibitemShut {NoStop}%
\bibitem [{\citenamefont {Lockwood}(2022)}]{lockwood2022empirical}%
  \BibitemOpen
  \bibfield  {author} {\bibinfo {author} {\bibfnamefont {O.}~\bibnamefont
  {Lockwood}},\ }\bibfield  {title} {\bibinfo {title} {An empirical review of
  optimization techniques for quantum variational circuits},\ }\href
  {https://arxiv.org/abs/2202.01389} {\bibfield  {journal} {\bibinfo  {journal}
  {arXiv preprint arXiv:2202.01389}\ } (\bibinfo {year} {2022})}\BibitemShut
  {NoStop}%
\bibitem [{\citenamefont {Wiersema}\ \emph {et~al.}(2020)\citenamefont
  {Wiersema}, \citenamefont {Zhou}, \citenamefont {de~Sereville}, \citenamefont
  {Carrasquilla}, \citenamefont {Kim},\ and\ \citenamefont
  {Yuen}}]{PRXQuantum.1.020319}%
  \BibitemOpen
  \bibfield  {author} {\bibinfo {author} {\bibfnamefont {R.}~\bibnamefont
  {Wiersema}}, \bibinfo {author} {\bibfnamefont {C.}~\bibnamefont {Zhou}},
  \bibinfo {author} {\bibfnamefont {Y.}~\bibnamefont {de~Sereville}}, \bibinfo
  {author} {\bibfnamefont {J.~F.}\ \bibnamefont {Carrasquilla}}, \bibinfo
  {author} {\bibfnamefont {Y.~B.}\ \bibnamefont {Kim}},\ and\ \bibinfo {author}
  {\bibfnamefont {H.}~\bibnamefont {Yuen}},\ }\bibfield  {title} {\bibinfo
  {title} {Exploring entanglement and optimization within the hamiltonian
  variational ansatz},\ }\href {https://doi.org/10.1103/PRXQuantum.1.020319}
  {\bibfield  {journal} {\bibinfo  {journal} {PRX Quantum}\ }\textbf {\bibinfo
  {volume} {1}},\ \bibinfo {pages} {020319} (\bibinfo {year}
  {2020})}\BibitemShut {NoStop}%
\bibitem [{\citenamefont {Anschuetz}(2021)}]{anschuetz2021critical}%
  \BibitemOpen
  \bibfield  {author} {\bibinfo {author} {\bibfnamefont {E.~R.}\ \bibnamefont
  {Anschuetz}},\ }\bibfield  {title} {\bibinfo {title} {Critical points in
  hamiltonian agnostic variational quantum algorithms},\ }\href
  {https://arxiv.org/abs/2109.06957} {\bibfield  {journal} {\bibinfo  {journal}
  {arXiv preprint arXiv:2109.06957}\ } (\bibinfo {year} {2021})}\BibitemShut
  {NoStop}%
\bibitem [{\citenamefont {Weidenfeller}\ \emph {et~al.}(2022)\citenamefont
  {Weidenfeller}, \citenamefont {Valor}, \citenamefont {Gacon}, \citenamefont
  {Tornow}, \citenamefont {Bello}, \citenamefont {Woerner},\ and\ \citenamefont
  {Egger}}]{weidenfeller2022scaling}%
  \BibitemOpen
  \bibfield  {author} {\bibinfo {author} {\bibfnamefont {J.}~\bibnamefont
  {Weidenfeller}}, \bibinfo {author} {\bibfnamefont {L.~C.}\ \bibnamefont
  {Valor}}, \bibinfo {author} {\bibfnamefont {J.}~\bibnamefont {Gacon}},
  \bibinfo {author} {\bibfnamefont {C.}~\bibnamefont {Tornow}}, \bibinfo
  {author} {\bibfnamefont {L.}~\bibnamefont {Bello}}, \bibinfo {author}
  {\bibfnamefont {S.}~\bibnamefont {Woerner}},\ and\ \bibinfo {author}
  {\bibfnamefont {D.~J.}\ \bibnamefont {Egger}},\ }\bibfield  {title} {\bibinfo
  {title} {Scaling of the quantum approximate optimization algorithm on
  superconducting qubit based hardware},\ }\href
  {https://arxiv.org/abs/2202.03459} {\bibfield  {journal} {\bibinfo  {journal}
  {arXiv preprint arXiv:2202.03459}\ } (\bibinfo {year} {2022})}\BibitemShut
  {NoStop}%
\bibitem [{\citenamefont {Zhu}\ \emph {et~al.}(2020)\citenamefont {Zhu},
  \citenamefont {Tang}, \citenamefont {Barron}, \citenamefont
  {Calderon-Vargas}, \citenamefont {Mayhall}, \citenamefont {Barnes},\ and\
  \citenamefont {Economou}}]{zhu2020adaptive}%
  \BibitemOpen
  \bibfield  {author} {\bibinfo {author} {\bibfnamefont {L.}~\bibnamefont
  {Zhu}}, \bibinfo {author} {\bibfnamefont {H.~L.}\ \bibnamefont {Tang}},
  \bibinfo {author} {\bibfnamefont {G.~S.}\ \bibnamefont {Barron}}, \bibinfo
  {author} {\bibfnamefont {F.}~\bibnamefont {Calderon-Vargas}}, \bibinfo
  {author} {\bibfnamefont {N.~J.}\ \bibnamefont {Mayhall}}, \bibinfo {author}
  {\bibfnamefont {E.}~\bibnamefont {Barnes}},\ and\ \bibinfo {author}
  {\bibfnamefont {S.~E.}\ \bibnamefont {Economou}},\ }\bibfield  {title}
  {\bibinfo {title} {An adaptive quantum approximate optimization algorithm for
  solving combinatorial problems on a quantum computer},\ }\href
  {https://arxiv.org/abs/2005.10258} {\bibfield  {journal} {\bibinfo  {journal}
  {arXiv preprint arXiv:2005.10258}\ } (\bibinfo {year} {2020})}\BibitemShut
  {NoStop}%
\bibitem [{\citenamefont {Hadfield}\ \emph {et~al.}(2019)\citenamefont
  {Hadfield}, \citenamefont {Wang}, \citenamefont {O’Gorman}, \citenamefont
  {Rieffel}, \citenamefont {Venturelli},\ and\ \citenamefont
  {Biswas}}]{a12020034}%
  \BibitemOpen
  \bibfield  {author} {\bibinfo {author} {\bibfnamefont {S.}~\bibnamefont
  {Hadfield}}, \bibinfo {author} {\bibfnamefont {Z.}~\bibnamefont {Wang}},
  \bibinfo {author} {\bibfnamefont {B.}~\bibnamefont {O’Gorman}}, \bibinfo
  {author} {\bibfnamefont {E.~G.}\ \bibnamefont {Rieffel}}, \bibinfo {author}
  {\bibfnamefont {D.}~\bibnamefont {Venturelli}},\ and\ \bibinfo {author}
  {\bibfnamefont {R.}~\bibnamefont {Biswas}},\ }\bibfield  {title} {\bibinfo
  {title} {From the quantum approximate optimization algorithm to a quantum
  alternating operator ansatz},\ }\href
  {https://www.mdpi.com/1999-4893/12/2/34} {\bibfield  {journal} {\bibinfo
  {journal} {Algorithms}\ }\textbf {\bibinfo {volume} {12}} (\bibinfo {year}
  {2019})}\BibitemShut {NoStop}%
\bibitem [{\citenamefont {Headley}\ \emph {et~al.}(2020)\citenamefont
  {Headley}, \citenamefont {M{\"u}ller}, \citenamefont {Martin}, \citenamefont
  {Solano}, \citenamefont {Sanz},\ and\ \citenamefont
  {Wilhelm}}]{headley2020approximating}%
  \BibitemOpen
  \bibfield  {author} {\bibinfo {author} {\bibfnamefont {D.}~\bibnamefont
  {Headley}}, \bibinfo {author} {\bibfnamefont {T.}~\bibnamefont {M{\"u}ller}},
  \bibinfo {author} {\bibfnamefont {A.}~\bibnamefont {Martin}}, \bibinfo
  {author} {\bibfnamefont {E.}~\bibnamefont {Solano}}, \bibinfo {author}
  {\bibfnamefont {M.}~\bibnamefont {Sanz}},\ and\ \bibinfo {author}
  {\bibfnamefont {F.~K.}\ \bibnamefont {Wilhelm}},\ }\bibfield  {title}
  {\bibinfo {title} {Approximating the quantum approximate optimisation
  algorithm},\ }\href {https://arxiv.org/abs/2002.12215} {\bibfield  {journal}
  {\bibinfo  {journal} {arXiv preprint arXiv:2002.12215}\ } (\bibinfo {year}
  {2020})}\BibitemShut {NoStop}%
\bibitem [{\citenamefont {Demirplak}\ and\ \citenamefont
  {Rice}(2003{\natexlab{b}})}]{Demirplak2003}%
  \BibitemOpen
  \bibfield  {author} {\bibinfo {author} {\bibfnamefont {M.}~\bibnamefont
  {Demirplak}}\ and\ \bibinfo {author} {\bibfnamefont {S.~A.}\ \bibnamefont
  {Rice}},\ }\bibfield  {title} {\bibinfo {title} {Adiabatic population
  transfer with control fields},\ }\href {https://doi.org/10.1021/jp030708a}
  {\bibfield  {journal} {\bibinfo  {journal} {The Journal of Physical Chemistry
  A}\ }\textbf {\bibinfo {volume} {107}},\ \bibinfo {pages} {9937} (\bibinfo
  {year} {2003}{\natexlab{b}})}\BibitemShut {NoStop}%
\bibitem [{\citenamefont {del Campo}\ \emph {et~al.}(2012)\citenamefont {del
  Campo}, \citenamefont {Rams},\ and\ \citenamefont
  {Zurek}}]{PhysRevLett.109.115703}%
  \BibitemOpen
  \bibfield  {author} {\bibinfo {author} {\bibfnamefont {A.}~\bibnamefont {del
  Campo}}, \bibinfo {author} {\bibfnamefont {M.~M.}\ \bibnamefont {Rams}},\
  and\ \bibinfo {author} {\bibfnamefont {W.~H.}\ \bibnamefont {Zurek}},\
  }\bibfield  {title} {\bibinfo {title} {Assisted finite-rate adiabatic passage
  across a quantum critical point: Exact solution for the quantum ising
  model},\ }\href {https://doi.org/10.1103/PhysRevLett.109.115703} {\bibfield
  {journal} {\bibinfo  {journal} {Physical Review Letters}\ }\textbf {\bibinfo
  {volume} {109}},\ \bibinfo {pages} {115703} (\bibinfo {year}
  {2012})}\BibitemShut {NoStop}%
\bibitem [{\citenamefont {Hospedales}\ \emph {et~al.}(2020)\citenamefont
  {Hospedales}, \citenamefont {Antoniou}, \citenamefont {Micaelli},\ and\
  \citenamefont {Storkey}}]{hospedales2021}%
  \BibitemOpen
  \bibfield  {author} {\bibinfo {author} {\bibfnamefont {T.}~\bibnamefont
  {Hospedales}}, \bibinfo {author} {\bibfnamefont {A.}~\bibnamefont
  {Antoniou}}, \bibinfo {author} {\bibfnamefont {P.}~\bibnamefont {Micaelli}},\
  and\ \bibinfo {author} {\bibfnamefont {A.}~\bibnamefont {Storkey}},\
  }\bibfield  {title} {\bibinfo {title} {Meta-learning in neural networks: A
  survey},\ }\href {https://arxiv.org/abs/2004.05439} {\bibfield  {journal}
  {\bibinfo  {journal} {arXiv preprint arXiv:2004.05439}\ } (\bibinfo {year}
  {2020})}\BibitemShut {NoStop}%
\bibitem [{\citenamefont {Cabessa}\ and\ \citenamefont
  {Siegelmann}(2012)}]{10.1162/NECO_a_00263}%
  \BibitemOpen
  \bibfield  {author} {\bibinfo {author} {\bibfnamefont {J.}~\bibnamefont
  {Cabessa}}\ and\ \bibinfo {author} {\bibfnamefont {H.~T.}\ \bibnamefont
  {Siegelmann}},\ }\bibfield  {title} {\bibinfo {title} {{The Computational
  Power of Interactive Recurrent Neural Networks}},\ }\href
  {https://doi.org/10.1162/NECO_a_00263} {\bibfield  {journal} {\bibinfo
  {journal} {Neural Computation}\ }\textbf {\bibinfo {volume} {24}},\ \bibinfo
  {pages} {996} (\bibinfo {year} {2012})}\BibitemShut {NoStop}%
\bibitem [{\citenamefont {Hochreiter}\ and\ \citenamefont
  {Schmidhuber}(1997)}]{hochreiter}%
  \BibitemOpen
  \bibfield  {author} {\bibinfo {author} {\bibfnamefont {S.}~\bibnamefont
  {Hochreiter}}\ and\ \bibinfo {author} {\bibfnamefont {J.}~\bibnamefont
  {Schmidhuber}},\ }\bibfield  {title} {\bibinfo {title} {Long short-term
  memory},\ }\href {https://doi.org/10.1162/neco.1997.9.8.1735} {\bibfield
  {journal} {\bibinfo  {journal} {Neural Computation}\ }\textbf {\bibinfo
  {volume} {9}},\ \bibinfo {pages} {1735} (\bibinfo {year} {1997})}\BibitemShut
  {NoStop}%
\bibitem [{\citenamefont {Cho}\ \emph {et~al.}(2014)\citenamefont {Cho},
  \citenamefont {Van~Merri{\"e}nboer}, \citenamefont {Bahdanau},\ and\
  \citenamefont {Bengio}}]{cho2014properties}%
  \BibitemOpen
  \bibfield  {author} {\bibinfo {author} {\bibfnamefont {K.}~\bibnamefont
  {Cho}}, \bibinfo {author} {\bibfnamefont {B.}~\bibnamefont
  {Van~Merri{\"e}nboer}}, \bibinfo {author} {\bibfnamefont {D.}~\bibnamefont
  {Bahdanau}},\ and\ \bibinfo {author} {\bibfnamefont {Y.}~\bibnamefont
  {Bengio}},\ }\bibfield  {title} {\bibinfo {title} {On the properties of
  neural machine translation: Encoder-decoder approaches},\ }\href
  {https://arxiv.org/abs/1409.1259} {\bibfield  {journal} {\bibinfo  {journal}
  {arXiv preprint arXiv:1409.1259}\ } (\bibinfo {year} {2014})}\BibitemShut
  {NoStop}%
\bibitem [{\citenamefont {Lechner}\ and\ \citenamefont
  {Hasani}(2020)}]{lechner2020learning}%
  \BibitemOpen
  \bibfield  {author} {\bibinfo {author} {\bibfnamefont {M.}~\bibnamefont
  {Lechner}}\ and\ \bibinfo {author} {\bibfnamefont {R.}~\bibnamefont
  {Hasani}},\ }\bibfield  {title} {\bibinfo {title} {Learning long-term
  dependencies in irregularly-sampled time series},\ }\href
  {https://arxiv.org/abs/2006.04418} {\bibfield  {journal} {\bibinfo  {journal}
  {arXiv preprint arXiv:2006.04418}\ } (\bibinfo {year} {2020})}\BibitemShut
  {NoStop}%
\bibitem [{\citenamefont {Noh}(2021)}]{info12110442}%
  \BibitemOpen
  \bibfield  {author} {\bibinfo {author} {\bibfnamefont {S.-H.}\ \bibnamefont
  {Noh}},\ }\bibfield  {title} {\bibinfo {title} {Analysis of gradient
  vanishing of rnns and performance comparison},\ }\href
  {https://www.mdpi.com/2078-2489/12/11/442} {\bibfield  {journal} {\bibinfo
  {journal} {Information}\ }\textbf {\bibinfo {volume} {12}} (\bibinfo {year}
  {2021})}\BibitemShut {NoStop}%
\bibitem [{\citenamefont {Verdon}\ \emph {et~al.}(2019)\citenamefont {Verdon},
  \citenamefont {Broughton}, \citenamefont {McClean}, \citenamefont {Sung},
  \citenamefont {Babbush}, \citenamefont {Jiang}, \citenamefont {Neven},\ and\
  \citenamefont {Mohseni}}]{verdon2019learning}%
  \BibitemOpen
  \bibfield  {author} {\bibinfo {author} {\bibfnamefont {G.}~\bibnamefont
  {Verdon}}, \bibinfo {author} {\bibfnamefont {M.}~\bibnamefont {Broughton}},
  \bibinfo {author} {\bibfnamefont {J.~R.}\ \bibnamefont {McClean}}, \bibinfo
  {author} {\bibfnamefont {K.~J.}\ \bibnamefont {Sung}}, \bibinfo {author}
  {\bibfnamefont {R.}~\bibnamefont {Babbush}}, \bibinfo {author} {\bibfnamefont
  {Z.}~\bibnamefont {Jiang}}, \bibinfo {author} {\bibfnamefont
  {H.}~\bibnamefont {Neven}},\ and\ \bibinfo {author} {\bibfnamefont
  {M.}~\bibnamefont {Mohseni}},\ }\bibfield  {title} {\bibinfo {title}
  {Learning to learn with quantum neural networks via classical neural
  networks},\ }\href {https://arxiv.org/abs/1907.05415} {\bibfield  {journal}
  {\bibinfo  {journal} {arXiv preprint arXiv:1907.05415}\ } (\bibinfo {year}
  {2019})}\BibitemShut {NoStop}%
\bibitem [{\citenamefont {Karp}(1972)}]{karp1972reducibility}%
  \BibitemOpen
  \bibfield  {author} {\bibinfo {author} {\bibfnamefont {R.~M.}\ \bibnamefont
  {Karp}},\ }\bibfield  {title} {\bibinfo {title} {Reducibility among
  combinatorial problems},\ }\href
  {https://www.semanticscholar.org/paper/Reducibility-Among-Combinatorial-Problems-Karp/9fb53a3bdfb47230eeaf7d956b1a238db5cba690}
  {\bibfield  {journal} {\bibinfo  {journal} {Complexity of computer
  computations}\ ,\ \bibinfo {pages} {85}} (\bibinfo {year}
  {1972})}\BibitemShut {NoStop}%
\bibitem [{\citenamefont {Guerreschi}\ and\ \citenamefont
  {Matsuura}(2019)}]{Guerreschi2019}%
  \BibitemOpen
  \bibfield  {author} {\bibinfo {author} {\bibfnamefont {G.~G.}\ \bibnamefont
  {Guerreschi}}\ and\ \bibinfo {author} {\bibfnamefont {A.~Y.}\ \bibnamefont
  {Matsuura}},\ }\bibfield  {title} {\bibinfo {title} {Qaoa for max-cut
  requires hundreds of qubits for quantum speed-up},\ }\href
  {https://doi.org/10.1038/s41598-019-43176-9} {\bibfield  {journal} {\bibinfo
  {journal} {Scientific Reports}\ }\textbf {\bibinfo {volume} {9}},\ \bibinfo
  {pages} {6903} (\bibinfo {year} {2019})}\BibitemShut {NoStop}%
\bibitem [{\citenamefont {Kingma}\ and\ \citenamefont
  {Ba}(2014)}]{kingma2014adam}%
  \BibitemOpen
  \bibfield  {author} {\bibinfo {author} {\bibfnamefont {D.~P.}\ \bibnamefont
  {Kingma}}\ and\ \bibinfo {author} {\bibfnamefont {J.}~\bibnamefont {Ba}},\
  }\bibfield  {title} {\bibinfo {title} {Adam: A method for stochastic
  optimization},\ }\href {https://arxiv.org/abs/1412.6980} {\bibfield
  {journal} {\bibinfo  {journal} {arXiv preprint arXiv:1412.6980}\ } (\bibinfo
  {year} {2014})}\BibitemShut {NoStop}%
\bibitem [{\citenamefont {Duchi}\ \emph {et~al.}(2011)\citenamefont {Duchi},
  \citenamefont {Hazan},\ and\ \citenamefont {Singer}}]{JMLR:v12:duchi11a}%
  \BibitemOpen
  \bibfield  {author} {\bibinfo {author} {\bibfnamefont {J.}~\bibnamefont
  {Duchi}}, \bibinfo {author} {\bibfnamefont {E.}~\bibnamefont {Hazan}},\ and\
  \bibinfo {author} {\bibfnamefont {Y.}~\bibnamefont {Singer}},\ }\bibfield
  {title} {\bibinfo {title} {Adaptive subgradient methods for online learning
  and stochastic optimization},\ }\href
  {http://jmlr.org/papers/v12/duchi11a.html} {\bibfield  {journal} {\bibinfo
  {journal} {Journal of Machine Learning Research}\ }\textbf {\bibinfo {volume}
  {12}},\ \bibinfo {pages} {2121} (\bibinfo {year} {2011})}\BibitemShut
  {NoStop}%
\bibitem [{\citenamefont {Shaydulin}\ \emph {et~al.}(2022)\citenamefont
  {Shaydulin}, \citenamefont {Lotshaw}, \citenamefont {Larson}, \citenamefont
  {Ostrowski},\ and\ \citenamefont {Humble}}]{shaydulin2022parameter}%
  \BibitemOpen
  \bibfield  {author} {\bibinfo {author} {\bibfnamefont {R.}~\bibnamefont
  {Shaydulin}}, \bibinfo {author} {\bibfnamefont {P.~C.}\ \bibnamefont
  {Lotshaw}}, \bibinfo {author} {\bibfnamefont {J.}~\bibnamefont {Larson}},
  \bibinfo {author} {\bibfnamefont {J.}~\bibnamefont {Ostrowski}},\ and\
  \bibinfo {author} {\bibfnamefont {T.~S.}\ \bibnamefont {Humble}},\ }\bibfield
   {title} {\bibinfo {title} {Parameter transfer for quantum approximate
  optimization of weighted maxcut},\ }\href {https://arxiv.org/abs/2201.11785}
  {\bibfield  {journal} {\bibinfo  {journal} {arXiv preprint arXiv:2201.11785}\
  } (\bibinfo {year} {2022})}\BibitemShut {NoStop}%
\bibitem [{\citenamefont {Brandao}\ \emph {et~al.}(2018)\citenamefont
  {Brandao}, \citenamefont {Broughton}, \citenamefont {Farhi}, \citenamefont
  {Gutmann},\ and\ \citenamefont {Neven}}]{brandao2018fixed}%
  \BibitemOpen
  \bibfield  {author} {\bibinfo {author} {\bibfnamefont {F.~G.}\ \bibnamefont
  {Brandao}}, \bibinfo {author} {\bibfnamefont {M.}~\bibnamefont {Broughton}},
  \bibinfo {author} {\bibfnamefont {E.}~\bibnamefont {Farhi}}, \bibinfo
  {author} {\bibfnamefont {S.}~\bibnamefont {Gutmann}},\ and\ \bibinfo {author}
  {\bibfnamefont {H.}~\bibnamefont {Neven}},\ }\bibfield  {title} {\bibinfo
  {title} {For fixed control parameters the quantum approximate optimization
  algorithm's objective function value concentrates for typical instances},\
  }\href {https://arxiv.org/abs/1812.04170} {\bibfield  {journal} {\bibinfo
  {journal} {arXiv preprint arXiv:1812.04170}\ } (\bibinfo {year}
  {2018})}\BibitemShut {NoStop}%
\bibitem [{\citenamefont {Akshay}\ \emph {et~al.}(2021)\citenamefont {Akshay},
  \citenamefont {Rabinovich}, \citenamefont {Campos},\ and\ \citenamefont
  {Biamonte}}]{PhysRevA.104.L010401}%
  \BibitemOpen
  \bibfield  {author} {\bibinfo {author} {\bibfnamefont {V.}~\bibnamefont
  {Akshay}}, \bibinfo {author} {\bibfnamefont {D.}~\bibnamefont {Rabinovich}},
  \bibinfo {author} {\bibfnamefont {E.}~\bibnamefont {Campos}},\ and\ \bibinfo
  {author} {\bibfnamefont {J.}~\bibnamefont {Biamonte}},\ }\bibfield  {title}
  {\bibinfo {title} {Parameter concentrations in quantum approximate
  optimization},\ }\href {https://doi.org/10.1103/PhysRevA.104.L010401}
  {\bibfield  {journal} {\bibinfo  {journal} {Physical Review A}\ }\textbf
  {\bibinfo {volume} {104}},\ \bibinfo {pages} {L010401} (\bibinfo {year}
  {2021})}\BibitemShut {NoStop}%
\bibitem [{\citenamefont {Mele}\ \emph {et~al.}(2022)\citenamefont {Mele},
  \citenamefont {Mbeng}, \citenamefont {Santoro}, \citenamefont {Collura},\
  and\ \citenamefont {Torta}}]{mele2022avoiding}%
  \BibitemOpen
  \bibfield  {author} {\bibinfo {author} {\bibfnamefont {A.~A.}\ \bibnamefont
  {Mele}}, \bibinfo {author} {\bibfnamefont {G.~B.}\ \bibnamefont {Mbeng}},
  \bibinfo {author} {\bibfnamefont {G.~E.}\ \bibnamefont {Santoro}}, \bibinfo
  {author} {\bibfnamefont {M.}~\bibnamefont {Collura}},\ and\ \bibinfo {author}
  {\bibfnamefont {P.}~\bibnamefont {Torta}},\ }\bibfield  {title} {\bibinfo
  {title} {Avoiding barren plateaus via transferability of smooth solutions in
  hamiltonian variational ansatz},\ }\href {https://arxiv.org/abs/2206.01982}
  {\bibfield  {journal} {\bibinfo  {journal} {arXiv preprint arXiv:2206.01982}\
  } (\bibinfo {year} {2022})}\BibitemShut {NoStop}%
\bibitem [{\citenamefont {Binder}\ and\ \citenamefont
  {Young}(1986)}]{RevModPhys.58.801}%
  \BibitemOpen
  \bibfield  {author} {\bibinfo {author} {\bibfnamefont {K.}~\bibnamefont
  {Binder}}\ and\ \bibinfo {author} {\bibfnamefont {A.~P.}\ \bibnamefont
  {Young}},\ }\bibfield  {title} {\bibinfo {title} {Spin glasses: Experimental
  facts, theoretical concepts, and open questions},\ }\href
  {https://doi.org/10.1103/RevModPhys.58.801} {\bibfield  {journal} {\bibinfo
  {journal} {Reviews of Modern Physics}\ }\textbf {\bibinfo {volume} {58}},\
  \bibinfo {pages} {801} (\bibinfo {year} {1986})}\BibitemShut {NoStop}%
\bibitem [{\citenamefont {Sherrington}\ and\ \citenamefont
  {Kirkpatrick}(1975)}]{PhysRevLett.35.1792}%
  \BibitemOpen
  \bibfield  {author} {\bibinfo {author} {\bibfnamefont {D.}~\bibnamefont
  {Sherrington}}\ and\ \bibinfo {author} {\bibfnamefont {S.}~\bibnamefont
  {Kirkpatrick}},\ }\bibfield  {title} {\bibinfo {title} {Solvable model of a
  spin-glass},\ }\href {https://doi.org/10.1103/PhysRevLett.35.1792} {\bibfield
   {journal} {\bibinfo  {journal} {Physical Review Letters}\ }\textbf {\bibinfo
  {volume} {35}},\ \bibinfo {pages} {1792} (\bibinfo {year}
  {1975})}\BibitemShut {NoStop}%
\bibitem [{\citenamefont {Lucas}(2014)}]{10.3389/fphy.2014.00005}%
  \BibitemOpen
  \bibfield  {author} {\bibinfo {author} {\bibfnamefont {A.}~\bibnamefont
  {Lucas}},\ }\bibfield  {title} {\bibinfo {title} {Ising formulations of many
  np problems},\ }\href
  {https://www.frontiersin.org/article/10.3389/fphy.2014.00005} {\bibfield
  {journal} {\bibinfo  {journal} {Frontiers in Physics}\ }\textbf {\bibinfo
  {volume} {2}} (\bibinfo {year} {2014})}\BibitemShut {NoStop}%
\bibitem [{\citenamefont {Harrigan}\ \emph {et~al.}(2021)\citenamefont
  {Harrigan}, \citenamefont {Sung}, \citenamefont {Neeley}, \citenamefont
  {Satzinger}, \citenamefont {Arute}, \citenamefont {Arya}, \citenamefont
  {Atalaya}, \citenamefont {Bardin}, \citenamefont {Barends}, \citenamefont
  {Boixo}, \citenamefont {Broughton}, \citenamefont {Buckley}, \citenamefont
  {Buell}, \citenamefont {Burkett}, \citenamefont {Bushnell}, \citenamefont
  {Chen}, \citenamefont {Chen}, \citenamefont {Chiaro}, \citenamefont
  {Collins}, \citenamefont {Courtney}, \citenamefont {Demura}, \citenamefont
  {Dunsworth}, \citenamefont {Eppens}, \citenamefont {Fowler}, \citenamefont
  {Foxen}, \citenamefont {Gidney}, \citenamefont {Giustina}, \citenamefont
  {Graff}, \citenamefont {Habegger}, \citenamefont {Ho}, \citenamefont {Hong},
  \citenamefont {Huang}, \citenamefont {Ioffe}, \citenamefont {Isakov},
  \citenamefont {Jeffrey}, \citenamefont {Jiang}, \citenamefont {Jones},
  \citenamefont {Kafri}, \citenamefont {Kechedzhi}, \citenamefont {Kelly},
  \citenamefont {Kim}, \citenamefont {Klimov}, \citenamefont {Korotkov},
  \citenamefont {Kostritsa}, \citenamefont {Landhuis}, \citenamefont {Laptev},
  \citenamefont {Lindmark}, \citenamefont {Leib}, \citenamefont {Martin},
  \citenamefont {Martinis}, \citenamefont {McClean}, \citenamefont {McEwen},
  \citenamefont {Megrant}, \citenamefont {Mi}, \citenamefont {Mohseni},
  \citenamefont {Mruczkiewicz}, \citenamefont {Mutus}, \citenamefont {Naaman},
  \citenamefont {Neill}, \citenamefont {Neukart}, \citenamefont {Niu},
  \citenamefont {O'Brien}, \citenamefont {O'Gorman}, \citenamefont {Ostby},
  \citenamefont {Petukhov}, \citenamefont {Putterman}, \citenamefont
  {Quintana}, \citenamefont {Roushan}, \citenamefont {Rubin}, \citenamefont
  {Sank}, \citenamefont {Skolik}, \citenamefont {Smelyanskiy}, \citenamefont
  {Strain}, \citenamefont {Streif}, \citenamefont {Szalay}, \citenamefont
  {Vainsencher}, \citenamefont {White}, \citenamefont {Yao}, \citenamefont
  {Yeh}, \citenamefont {Zalcman}, \citenamefont {Zhou}, \citenamefont {Neven},
  \citenamefont {Bacon}, \citenamefont {Lucero}, \citenamefont {Farhi},\ and\
  \citenamefont {Babbush}}]{Harrigan2021}%
  \BibitemOpen
  \bibfield  {author} {\bibinfo {author} {\bibfnamefont {M.~P.}\ \bibnamefont
  {Harrigan}}, \bibinfo {author} {\bibfnamefont {K.~J.}\ \bibnamefont {Sung}},
  \bibinfo {author} {\bibfnamefont {M.}~\bibnamefont {Neeley}}, \bibinfo
  {author} {\bibfnamefont {K.~J.}\ \bibnamefont {Satzinger}}, \bibinfo {author}
  {\bibfnamefont {F.}~\bibnamefont {Arute}}, \bibinfo {author} {\bibfnamefont
  {K.}~\bibnamefont {Arya}}, \bibinfo {author} {\bibfnamefont {J.}~\bibnamefont
  {Atalaya}}, \bibinfo {author} {\bibfnamefont {J.~C.}\ \bibnamefont {Bardin}},
  \bibinfo {author} {\bibfnamefont {R.}~\bibnamefont {Barends}}, \bibinfo
  {author} {\bibfnamefont {S.}~\bibnamefont {Boixo}}, \bibinfo {author}
  {\bibfnamefont {M.}~\bibnamefont {Broughton}}, \bibinfo {author}
  {\bibfnamefont {B.~B.}\ \bibnamefont {Buckley}}, \bibinfo {author}
  {\bibfnamefont {D.~A.}\ \bibnamefont {Buell}}, \bibinfo {author}
  {\bibfnamefont {B.}~\bibnamefont {Burkett}}, \bibinfo {author} {\bibfnamefont
  {N.}~\bibnamefont {Bushnell}}, \bibinfo {author} {\bibfnamefont
  {Y.}~\bibnamefont {Chen}}, \bibinfo {author} {\bibfnamefont {Z.}~\bibnamefont
  {Chen}}, \bibinfo {author} {\bibfnamefont {B.}~\bibnamefont {Chiaro}},
  \bibinfo {author} {\bibfnamefont {R.}~\bibnamefont {Collins}}, \bibinfo
  {author} {\bibfnamefont {W.}~\bibnamefont {Courtney}}, \bibinfo {author}
  {\bibfnamefont {S.}~\bibnamefont {Demura}}, \bibinfo {author} {\bibfnamefont
  {A.}~\bibnamefont {Dunsworth}}, \bibinfo {author} {\bibfnamefont
  {D.}~\bibnamefont {Eppens}}, \bibinfo {author} {\bibfnamefont
  {A.}~\bibnamefont {Fowler}}, \bibinfo {author} {\bibfnamefont
  {B.}~\bibnamefont {Foxen}}, \bibinfo {author} {\bibfnamefont
  {C.}~\bibnamefont {Gidney}}, \bibinfo {author} {\bibfnamefont
  {M.}~\bibnamefont {Giustina}}, \bibinfo {author} {\bibfnamefont
  {R.}~\bibnamefont {Graff}}, \bibinfo {author} {\bibfnamefont
  {S.}~\bibnamefont {Habegger}}, \bibinfo {author} {\bibfnamefont
  {A.}~\bibnamefont {Ho}}, \bibinfo {author} {\bibfnamefont {S.}~\bibnamefont
  {Hong}}, \bibinfo {author} {\bibfnamefont {T.}~\bibnamefont {Huang}},
  \bibinfo {author} {\bibfnamefont {L.~B.}\ \bibnamefont {Ioffe}}, \bibinfo
  {author} {\bibfnamefont {S.~V.}\ \bibnamefont {Isakov}}, \bibinfo {author}
  {\bibfnamefont {E.}~\bibnamefont {Jeffrey}}, \bibinfo {author} {\bibfnamefont
  {Z.}~\bibnamefont {Jiang}}, \bibinfo {author} {\bibfnamefont
  {C.}~\bibnamefont {Jones}}, \bibinfo {author} {\bibfnamefont
  {D.}~\bibnamefont {Kafri}}, \bibinfo {author} {\bibfnamefont
  {K.}~\bibnamefont {Kechedzhi}}, \bibinfo {author} {\bibfnamefont
  {J.}~\bibnamefont {Kelly}}, \bibinfo {author} {\bibfnamefont
  {S.}~\bibnamefont {Kim}}, \bibinfo {author} {\bibfnamefont {P.~V.}\
  \bibnamefont {Klimov}}, \bibinfo {author} {\bibfnamefont {A.~N.}\
  \bibnamefont {Korotkov}}, \bibinfo {author} {\bibfnamefont {F.}~\bibnamefont
  {Kostritsa}}, \bibinfo {author} {\bibfnamefont {D.}~\bibnamefont {Landhuis}},
  \bibinfo {author} {\bibfnamefont {P.}~\bibnamefont {Laptev}}, \bibinfo
  {author} {\bibfnamefont {M.}~\bibnamefont {Lindmark}}, \bibinfo {author}
  {\bibfnamefont {M.}~\bibnamefont {Leib}}, \bibinfo {author} {\bibfnamefont
  {O.}~\bibnamefont {Martin}}, \bibinfo {author} {\bibfnamefont {J.~M.}\
  \bibnamefont {Martinis}}, \bibinfo {author} {\bibfnamefont {J.~R.}\
  \bibnamefont {McClean}}, \bibinfo {author} {\bibfnamefont {M.}~\bibnamefont
  {McEwen}}, \bibinfo {author} {\bibfnamefont {A.}~\bibnamefont {Megrant}},
  \bibinfo {author} {\bibfnamefont {X.}~\bibnamefont {Mi}}, \bibinfo {author}
  {\bibfnamefont {M.}~\bibnamefont {Mohseni}}, \bibinfo {author} {\bibfnamefont
  {W.}~\bibnamefont {Mruczkiewicz}}, \bibinfo {author} {\bibfnamefont
  {J.}~\bibnamefont {Mutus}}, \bibinfo {author} {\bibfnamefont
  {O.}~\bibnamefont {Naaman}}, \bibinfo {author} {\bibfnamefont
  {C.}~\bibnamefont {Neill}}, \bibinfo {author} {\bibfnamefont
  {F.}~\bibnamefont {Neukart}}, \bibinfo {author} {\bibfnamefont {M.~Y.}\
  \bibnamefont {Niu}}, \bibinfo {author} {\bibfnamefont {T.~E.}\ \bibnamefont
  {O'Brien}}, \bibinfo {author} {\bibfnamefont {B.}~\bibnamefont {O'Gorman}},
  \bibinfo {author} {\bibfnamefont {E.}~\bibnamefont {Ostby}}, \bibinfo
  {author} {\bibfnamefont {A.}~\bibnamefont {Petukhov}}, \bibinfo {author}
  {\bibfnamefont {H.}~\bibnamefont {Putterman}}, \bibinfo {author}
  {\bibfnamefont {C.}~\bibnamefont {Quintana}}, \bibinfo {author}
  {\bibfnamefont {P.}~\bibnamefont {Roushan}}, \bibinfo {author} {\bibfnamefont
  {N.~C.}\ \bibnamefont {Rubin}}, \bibinfo {author} {\bibfnamefont
  {D.}~\bibnamefont {Sank}}, \bibinfo {author} {\bibfnamefont {A.}~\bibnamefont
  {Skolik}}, \bibinfo {author} {\bibfnamefont {V.}~\bibnamefont {Smelyanskiy}},
  \bibinfo {author} {\bibfnamefont {D.}~\bibnamefont {Strain}}, \bibinfo
  {author} {\bibfnamefont {M.}~\bibnamefont {Streif}}, \bibinfo {author}
  {\bibfnamefont {M.}~\bibnamefont {Szalay}}, \bibinfo {author} {\bibfnamefont
  {A.}~\bibnamefont {Vainsencher}}, \bibinfo {author} {\bibfnamefont
  {T.}~\bibnamefont {White}}, \bibinfo {author} {\bibfnamefont {Z.~J.}\
  \bibnamefont {Yao}}, \bibinfo {author} {\bibfnamefont {P.}~\bibnamefont
  {Yeh}}, \bibinfo {author} {\bibfnamefont {A.}~\bibnamefont {Zalcman}},
  \bibinfo {author} {\bibfnamefont {L.}~\bibnamefont {Zhou}}, \bibinfo {author}
  {\bibfnamefont {H.}~\bibnamefont {Neven}}, \bibinfo {author} {\bibfnamefont
  {D.}~\bibnamefont {Bacon}}, \bibinfo {author} {\bibfnamefont
  {E.}~\bibnamefont {Lucero}}, \bibinfo {author} {\bibfnamefont
  {E.}~\bibnamefont {Farhi}},\ and\ \bibinfo {author} {\bibfnamefont
  {R.}~\bibnamefont {Babbush}},\ }\bibfield  {title} {\bibinfo {title} {Quantum
  approximate optimization of non-planar graph problems on a planar
  superconducting processor},\ }\href
  {https://doi.org/10.1038/s41567-020-01105-y} {\bibfield  {journal} {\bibinfo
  {journal} {Nature Physics}\ }\textbf {\bibinfo {volume} {17}},\ \bibinfo
  {pages} {332} (\bibinfo {year} {2021})}\BibitemShut {NoStop}%
\bibitem [{\citenamefont {Farhi}\ \emph {et~al.}(2019)\citenamefont {Farhi},
  \citenamefont {Goldstone}, \citenamefont {Gutmann},\ and\ \citenamefont
  {Zhou}}]{farhi2019quantum2}%
  \BibitemOpen
  \bibfield  {author} {\bibinfo {author} {\bibfnamefont {E.}~\bibnamefont
  {Farhi}}, \bibinfo {author} {\bibfnamefont {J.}~\bibnamefont {Goldstone}},
  \bibinfo {author} {\bibfnamefont {S.}~\bibnamefont {Gutmann}},\ and\ \bibinfo
  {author} {\bibfnamefont {L.}~\bibnamefont {Zhou}},\ }\bibfield  {title}
  {\bibinfo {title} {The quantum approximate optimization algorithm and the
  sherrington-kirkpatrick model at infinite size},\ }\href
  {https://arxiv.org/abs/1910.08187} {\bibfield  {journal} {\bibinfo  {journal}
  {arXiv preprint arXiv:1910.08187}\ } (\bibinfo {year} {2019})}\BibitemShut
  {NoStop}%
\end{thebibliography}%

\end{document}